\documentclass[11pt,a4paper]{article}

\usepackage[T2A]{fontenc}
\usepackage[english]{babel}

\usepackage[
  a4paper,
  textwidth=17cm,
  textheight=23cm,
  centering
]{geometry}

\usepackage{
  amsmath,
  amssymb,
  amsfonts,
  amsthm,
  mathtools,
  bm,
  braket,
  slashed
}
\usepackage[makeroom]{cancel}

\usepackage{
  booktabs,
  multirow,
  tabularx,
  arydshln
}

\usepackage{graphicx}
\usepackage{float}
\usepackage[pdftex]{pict2e}

\usepackage[
  usenames,
  dvipsnames,
  svgnames,
  table
]{xcolor}

\usepackage{tikz}
\usetikzlibrary{
  positioning,
  fit,
  calc,
  arrows.meta,
  decorations.pathreplacing
}

\usepackage{pgfplots}
\pgfplotsset{compat=1.18}
\usepgfplotslibrary{groupplots}

\usepackage{titlesec}
\titleformat{\section}
  {\normalfont\Large\bfseries}
  {\thesection}{1em}{}

\usepackage{tcolorbox}
\usepackage{textcomp}
\usepackage{marvosym}

\usepackage{cite}
\usepackage{verbatim}

\usepackage[
  pageanchor=false,
  colorlinks=true,
  linktocpage=true,
  breaklinks=true,
  urlcolor=blue,
  linkcolor=blue,
  citecolor=blue
]{hyperref}

\numberwithin{equation}{section}

\newcommand{\NC}{\ket{\mathrm{NC}}}

\newcommand{\CNOT}{\mathrm{CNOT}}
\newcommand{\Bsix}{B_6}
\DeclareMathOperator{\Tr}{Tr}

\newtheoremstyle{strongtheorem}
  {1em}
  {1em}
  {\normalfont}
  {}
  {\normalfont\bfseries}
  {.}
  {0.6em}
  {\thmname{#1}\thmnumber{ #2}\thmnote{ — #3}}

\theoremstyle{strongtheorem}

\newtheorem{theorem}{\iflanguage{russian}{Теорема}{Theorem}}[section]
\newtheorem{lemma}[theorem]{\iflanguage{russian}{Лемма}{Lemma}}
\newtheorem{corollary}[theorem]{\iflanguage{russian}{Следствие}{Corollary}}
\newtheorem{definition}[theorem]{Definition}

\tikzset{
  modern/.style={
    rectangle,
    rounded corners=8pt,
    minimum width=3.5cm,
    minimum height=1.2cm,
    text centered,
    align=center,
    font=\sffamily\bfseries,
    fill=#1!15,
    draw=#1!80!black
  },
  arrowstyle/.style={
    -{Stealth[length=3mm,width=2mm]},
    thick,
    draw=black!85,
    rounded corners=8pt
  }
}

\def\newanyon(#1,#2){%
  \draw[blue,thick]
    (#1,#2) .. controls (#1,#2-0.8)
    and (#1+1,#2-0.8) .. (#1+1,#2);
}

\def\anyonend(#1,#2){%
  \draw[blue,thick]
    (#1,#2) .. controls (#1+0.05,#2+0.4)
    and (#1+0.45,#2+0.4) .. (#1+0.5,#2);
}

\def\measureup(#1,#2){%
  \draw[thick,blue]
    (#1,#2+0.5) .. controls (#1+0.05,#2+0.1)
    and (#1+0.45,#2+0.1) .. (#1+0.5,#2+0.5);
  \draw[thick,blue]
    (#1+0.25,#2+0.2) -- (#1+0.25,#2);
}

\def\measuredown(#1,#2){%
  \draw[thick,blue]
    (#1,#2) .. controls (#1+0.05,#2+0.4)
    and (#1+0.45,#2+0.4) .. (#1+0.5,#2);
  \draw[thick,blue]
    (#1+0.25,#2+0.3) -- (#1+0.25,#2+0.5);
}

\def\rcross(#1,#2){%
  \draw[
    draw=white,
    double=blue,
    ultra thick,
    double distance=0.8pt
  ]
    (#1,#2) .. controls (#1,#2+0.5)
    and (#1+1,#2+0.5) .. (#1+1,#2+1);

  \draw[
    draw=white,
    double=blue,
    ultra thick,
    double distance=0.8pt
  ]
    (#1+1,#2) .. controls (#1+1,#2+0.5)
    and (#1,#2+0.5) .. (#1,#2+1);
}

\def\lcross(#1,#2){%
  \draw[
    draw=white,
    double=blue,
    ultra thick,
    double distance=0.8pt
  ]
    (#1+1,#2) .. controls (#1+1,#2+0.5)
    and (#1,#2+0.5) .. (#1,#2+1);

  \draw[
    draw=white,
    double=blue,
    ultra thick,
    double distance=0.8pt
  ]
    (#1,#2) .. controls (#1,#2+0.5)
    and (#1+1,#2+0.5) .. (#1+1,#2+1);
}

\title{%
  \parbox{\textwidth}{%
    \normalfont\normalsize\raggedleft
    ITEP/TH-39/26\\
    IITP/TH-33/26\\
    MIPT/TH-30/26
  }
  \par\vspace{1em}
  \bfseries
  Meet-in-the-Middle Compilation of the $[\mathrm{CNOT}]$ Class
  with Fibonacci Anyons
}

\author{%
  Matvei Fedin$^{a,b,c}$%
  \thanks{E-mail: \texttt{fedin.mm@iitp.ru},
  \texttt{fedin.mm@phystech.edu}}
  \qquad
  Andrey Morozov%
  \thanks{E-mail: \texttt{morozov.andrey.a@iitp.ru}}
  \\[0.6em]
  \small $^{a}$Moscow Institute of Physics and Technology,
  141702 Dolgoprudny, Russia
  \\
  \small $^{b}$Institute for Information Transmission Problems,
  127051 Moscow, Russia
  \\
  \small $^{c}$ITMO University,
  197101 St.\ Petersburg, Russia
  \\
  \small $^{d}$ITEP, Moscow, Russia
}

\date{}

\begin{document}

\maketitle

\begin{abstract}

Compilation of two-qubit Fibonacci anyon gates is complicated by leakage ($\ell$) from the four-dimensional computational sector to the fifth state of the fusion space. We construct a GPU pipeline to find low-leakage representatives of the $[\CNOT]$ class on different subsets of the braid group.

We exhaustively explore the full positive five-generator alphabet up to $L=21$ with minimal leakage $\ell\sim10^{-3}$ and error $d_{[\CNOT]}\sim10^{-8}$ on it. To reach larger lengths, the search space is successively narrowed to three-generator, boundary-$\sigma_3$, and palindromic braids, and by considering only the Garside left normal form. These optimizations allow us to reach $L=72$. For $L=69$, a braid with $\ell\sim10^{-10}$ and $d_{[\CNOT]}\sim10^{-10}$ is found.

In addition to directly enumerating optimal braids, we proved that $\ell\in\mathbb Z[\varphi]$, and any nonzero leakage obeys an exponential lower bound in the braid length, and obtained the limit laws $F_{3,\infty}(\ell)=\ell^2$ and $F_{5,\infty}(\ell)=\ell^4$ for our sets. Using asymptotics for the distribution function $F$, we showed that for sufficiently long braids, the asymptotics of exhaustive search are better than the asymptotics of random sampling for finding low-leakage braids.

\end{abstract}

\newpage
\tableofcontents{}
\newpage

\section{Introduction}

Quantum computing is one of the most promising complements to classical computing. Grover's algorithm performs unstructured search in $O(\sqrt{N})$ instead of $O(N)$~\cite{Grover1996}, Shor's algorithm solves factorization and discrete logarithm problems in polynomial time, while no classical polynomial algorithms are known~\cite{Shor1997,GidneyEkera2021}. In some cases, exponential complexity is replaced by polynomial~\cite{Simon1997}. However, the global quantum advantage cannot be considered a proven conjecture~\cite{BernsteinVazirani1997}. Furthermore, comparable classical algorithms were later constructed for individual quantum algorithms, in particular QML (quantum machine learning) algorithms~\cite{Tang2019, ChiaEtAl2020, IsogaiMuraoYamasaki2026}. While probabilistic classical computations can be efficiently simulated on a quantum computer, no general efficient classical algorithm for simulating arbitrary quantum circuits is known~\cite{Bennett1973, BernsteinVazirani1997}. Therefore, it is natural to view a quantum processor as a specialized coprocessor for certain classes of problems~\cite{BeverlandEtAl2022}. The most obvious and natural problem solved by a quantum computer is the simulation of a quantum system, which is often exponentially complex for a classical computer and polynomial for a quantum one~\cite{Feynman1982,Lloyd1996,BerryAhokasCleveSanders2007,LowChuang2019}.

While quantum computing is a rapidly developing technology, there are many architectures for its implementation -- physical implementations of logical quantum computations. Each such architecture is defined by a (usually finite) set of base matrices, generators, whose products approximate the target unitary transformations. A set is called universal if the operations it generates are dense in the corresponding unitary group up to an overall phase~\cite{DawsonNielsen2006,BrylinskiBrylinski2002}.

Physical qubits are susceptible to decoherence through energy relaxation and dephasing, as well as to control, crosstalk, state-preparation, measurement, and leakage errors~\cite{NielsenChuang2010,Krantz2019}. This problem, in a broad sense, also exists in classical computing and is not exclusively an artifact of the different physics underlying a quantum processor. In both classical and quantum computing, physical errors can be suppressed through redundant encoding and recovery of information~\cite{Hamming1950,vonNeumann1956,ShorQEC1995,AharonovBenOr2008}. However, the solution to this problem in the quantum case is complicated by quantum-mechanical features, such as the fundamental impossibility of copying an unknown quantum state and the need to detect errors without measuring the logical information itself~\cite{WoottersZurek1982,NielsenChuang2010}. One approach, Surface Code~\cite{DennisKitaevLandahlPreskill2002,FowlerEtAl2012,Litinski2019}, encodes a single qubit as a set of physical qubits, and the number of physical qubits in a single surface-code patch scales as $O(d^2)$ and can be hundreds or thousands, depending on the code distance, physical error, and required logical error. Alternative families include quantum LDPC codes, topological color codes, and bosonic codes~\cite{BreuckmannEberhardt2021,BombinMartinDelgado2006,GKP2001}.

In this paper, we discuss a different approach -- a topological quantum computer~\cite{Kitaev2003,NayakRMP2008}. Quantum information is encoded in the degenerate fusion space of non-Abelian anyons, while logical gates are implemented by adiabatically exchanging the anyons. These exchanges generate the braid group, whose generators are represented by unitary matrices acting on the fusion space. Since the resulting transformation depends only on the topological class of the world-line braid, it is insensitive to sufficiently small local deformations of the particle trajectories~\cite{Kitaev2003,NayakRMP2008}. Therefore, errors caused by such local perturbations are expected to be suppressed by construction.

Even within the class of topological quantum computers, there are various anyon models generating different representations of the braid group, including the Ising anyons in $SU(2)_2$~\cite{BelovMorozovIsing2026,Bravyi2006}, the $SU(2)_k$ and $SU(N)_k$~\cite{KolganovMorozov2020, KolganovMironovMorozov2023,BelovMorozov2024} families, and the Fibonacci anyons~\cite{NayakRMP2008,FreedmanLarsenWang2002Universal}. The gate construction, topological parameter measurements, and two-qubit cable constructions for these theories have been studied in~\cite{MorozovTopologicalCharge2024,Cabling2025,MironovMorozov2026}.

In this paper, we consider a two-qubit TQC, which is encoded by six Fibonacci anyons. It defines a five-dimensional representation of the group $B_6$ with five positive generators $\sigma_1,\ldots,\sigma_5$ and five inverses $\sigma_1^{-1},\ldots,\sigma_5^{-1}$~\cite{CuiWang2019}.

The primary interest in using a quantum computer arises only under the condition of interaction between logical qubits. Independent single-qubit operations by themselves do not implement a universal multi-qubit computation. Arbitrary single-qubit operations together with any entangling two-qubit gate, on the contrary, form a universal set \cite{BrylinskiBrylinski2002,BremnerEtAl2002}. For pure-state quantum computations, states that remain decomposable into entangled blocks of bounded size, or more generally have polynomially bounded Schmidt rank across the relevant bipartitions, can be simulated efficiently on a classical computer~\cite{JozsaLinden2003,Vidal2003}. In this setting, we are interested in finding a representation of a special class $[\CNOT]$ in the architecture of interest. $[\CNOT]$ is a class of operators that are reducible by single-qubit operations to the matrix $\begin{bmatrix} 1&0&0&0\\0&1&0&0\\0&0&0&1\\0&0&1&0\\
\end{bmatrix}$. All representatives of this class are entangling, and CNOT together with arbitrary single-qubit operations forms a universal set \cite{Makhlin2002,BarencoEtAl1995}.

While the efficient search for a given-precision factorization for a single-qubit operator has been very well studied~\cite{HormoziEtAl2007,KBS2014}, the search for two-qubit operators faces various unexpected difficulties, especially in anyon theories. One of the key challenges in building a topological quantum computer is the mismatch between the dimensions of the fusion spaces of anyons and qubits. For example, in the case of Fibonacci anyons, the fusion spaces have dimensions $1,1,2,3,5,\dots$, while the state spaces of the qubits have the sequence $2,4,8,\dots$. In this case, to simulate a two-qubit gate, it is necessary to reduce the influence of the additional fifth dimension. In practice, this is achieved by trying to make the amplitude modulus $|U_{00}|$ approach unity. The quality of this approach to unity is measured through the leakage $\ell$:

\begin{equation}
\ell(U) = 1 - |U_{00}|^2
\label{eq:leak_intro}
\end{equation}

The goal of such optimization is to simultaneously make the distance to the target class $[\CNOT]$ and the leakage $\ell$ tend to zero. Most papers in this area rely on stochastic methods and iterative improvements, while methods with guarantees in a given search space are used less frequently due to the exponentially increasing computational complexity of the problem.

In some papers, special low-leakage and systematically generated two-qubit Fibonacci braids were constructed~\cite{XuWan2008,Carnahan2016}. Other directions include the single-qubit Fibonacci compilation~\cite{Long2025Fib}, genetic optimization for metaplectic and other $SU(2)_k$ models~\cite{Metaplectic2025,Long2026b}, multiple-braiding constructions~\cite{Long2026a}, non-semisimple Ising models~\cite{Long2025NSI}, and mixed-integer compilation~\cite{Rytir2025}. These works use different models, alphabets, or objective functions, so their lengths cannot be directly compared to the lengths of positive words in the elementary $\sigma_i$ considered here.

Our approach and the results presented in this paper belong to the class of exhaustive methods. One of the main works on exhaustive search for braids lying in a nearby neighborhood $[\CNOT]$ is due to Burke et al.~\cite{Burke2024} and in many ways serves as a benchmark for our work, which is largely inspired by it. Following the basic-alphabet setting of Burke et al.~\cite{Burke2024}, we restrict the search to $5$ positive generators. In~\cite{Burke2024}, an extended ten-letter alphabet with inverse generators did not provide a significant improvement on the available computational horizons, while a smaller alphabet allowed us to extend the exhaustive search to a larger length. This approach reduces the naive asymptotics of the number of braids of a given length $L$ from $10^L$ to $5^L$. We consistently improve the asymptotics of exhaustive search by gradually narrowing the search space in a motivated manner:

\begin{equation}
    5^L \longrightarrow 3^L \longrightarrow 3^{L-2} \longrightarrow 3^{L/2-2} \longrightarrow (2.086791\ldots)^{L/2}.
    \label{eq:opt_chain}
\end{equation}

We also use an algorithm inspired by the Meet-In-The-Middle (MITM) method~\cite{AmyMITM2013}, which allows us to store and reuse a portion of the intermediate computations. Together, this allows us to successively extend the horizon of exhaustively explored braids from the length $L=14$ in Burke et al.~\cite{Burke2024} to $L=72$. Despite the narrowing of the search space, the empirical dependence of the upper bound on leakage as a function of braid length remains exponential with the same coefficient on all sets and is close to the dependence $\log_{10} \ell_L = -\frac{L}{6}+O(1)$.

Furthermore, we studied the distribution functions of braids over leakages depending on the set under study and compared the asymptotic behavior of exhaustive search and random sampling on exhaustive sets and horizons. We concluded that random sampling has complexity no better than $\Theta(\ell^{-2})$, while exhaustive search is no worse than $\Theta(\ell^{-3\log_{10}3})$. Given that $\ell$ tends to $0$, this indicates that random sampling is pointless to use in this work for $L \sim 70$ that we consider.

The article is organized as follows. Section~\ref{sec:background} introduces Fibonacci anyons, six-anyons encoding, leakage, and the local class $[\CNOT]$. Section~\ref{sec:setup} describes the search space, MITM-GEMM (\emph{GEneral Matrix-Multiplication}) pipeline and optimization criteria. Section~\ref{sec:results} provides the obtained frontiers, structural reductions, and comparison with sampling. Appendix~\ref{app:fibonacci} derives the dimensions of the Fibonacci fusion spaces. Appendix~\ref{app:leak_CDF} proves the limiting leakage distribution laws.

\subsection{Software and hardware}\label{sec:bg-soft}

The results of this paper are the product of joint improvements to both the algorithmic and hardware sides of $[\CNOT]$ class searching methods. All computations were performed on a high-performance GPU node. The enumeration and GEMM are executed in \texttt{complex64}; the metrics of published candidates are recalculated in \texttt{float128}; and algebraic leakage assertions are verified in exact arithmetic. The implementation is split into resumable chunks: after each chunk, the progress, the current best candidate, and the launch parameters are saved, so long experiments can be safely interrupted and continued. The computational cost of individual points could reach $200$ GPU hours. The total number of GPU hours for a single V100 graphics card is approximately $1000$. On $4$ V100s, this required approximately $250$ hours of continuous operation. The specific Software and Hardware stack are given in Tab.~\ref{tab:software} and Tab.~\ref{tab:hardware}.

\begin{table}[htbp]
\centering
\footnotesize
\begin{minipage}[t]{0.52\textwidth}
  \centering
  \begin{tabularx}{\linewidth}{@{}l >{\raggedright\arraybackslash}X@{}}
  \hline
  Component & Version \\
  \hline
  OS & Debian 12 (Bookworm)\\
  Python & 3.11 \\
  PyTorch / torchvision & 2.4.1 / 0.19.1 (CUDA 12.1) \\
  CUDA runtime / compiler & 12.1 / 11.8 (nvcc) \\

  \hline
  \end{tabularx}
  \caption{Software stack.}
  \label{tab:software}
\end{minipage}\hfill
\begin{minipage}[t]{0.46\textwidth}
  \centering
  \begin{tabularx}{\linewidth}{@{}l >{\raggedright\arraybackslash}X@{}}
  \hline
  Component & Specification \\
  \hline
  CPU & AMD EPYC 7F52 (16 cores) \\
  RAM & 64\,GB \\
  GPU & $4 \times$NVIDIA Tesla V100-SXM2 \\
  VRAM &128\,GB\\
  \hline
  \end{tabularx}
  \caption{Hardware stack.}
  \label{tab:hardware}
\end{minipage}
\end{table}

\section{Introduction into quantum computation}\label{sec:background}

Before moving on to the main part of this work, let's review the basic concepts of topological quantum computing. More details about them can be found in~\cite{Kitaev2003,NayakRMP2008}.

\subsection{Quantum gates and the decoherence problem}

A quantum computer program is a unitary matrix acting on the state space of $n$ qubits, $U\in U(2^n)$. As in the classical case, the hardware can only execute a fixed set of elementary operations --- \emph{gates}. Any operator must be decomposed into a product of these basic gates. The main problem with such a decomposition is decoherence: uncontrolled interactions between qubits and their environment destroy the quantum superposition within a characteristic time, and each additional gate in the circuit increases the accumulated error. Typically, error-correcting codes are used to suppress this effect. They correct errors within the computational space at the cost of redundancy in the number of physical qubits. A single logical qubit can be realized by hundreds or thousands of physical qubits~\cite{Ruiz2025,GidneyYoked2025,KlimovEtAl2024}.

The topological approach~\cite{Kitaev2003} offers a different solution to this problem: information is encoded in the system's global, non-local degrees of freedom. Such a system is, in principle, insensitive to local noise. One proposed realization uses systems of non-Abelian anyons, with quantum information encoded in their collective fusion states~\cite{Kitaev2003,NayakRMP2008}.

\subsection{CNOT and the local equivalence class \texorpdfstring{$[\CNOT]$}{[CNOT]} \label{sec:makhlin_inv}}

Since single-qubit rotations are compiled separately and relatively cheaply~\cite{KBS2014}, we are interested in building $2$-qubit gates. The target is not a single fixed matrix $\CNOT$, but its entire local equivalence class:

\begin{equation}
[\CNOT]=\{(V_1\otimes V_2)\,\CNOT\,(V_3\otimes V_4):V_i\in U(2)\}.
\label{eq:CNOT_class}
\end{equation}

Elements of the same class differ only in local operations before and after the entangling gate and are therefore interchangeable in a two-qubit circuit, provided the surrounding single-qubit gates are adjusted accordingly. To recognize an equivalence class, we use invariants that do not change under local operations $V_i\in U(2)$. In particular, we use the Makhlin invariants~\cite{Makhlin2002}. The magic basis is a phase-adjusted Bell basis in which local $SU(2)\otimes SU(2)$ operations are represented by real orthogonal matrices. The columns of $Q$ are the magic-basis vectors expressed in the computational basis:

\begin{equation}
Q=\frac{1}{\sqrt{2}}
\begin{pmatrix}
1&0&0&i\\
0&i&1&0\\
0&i&-1&0\\
1&0&0&-i
\end{pmatrix}.
\end{equation}

For a two-qubit matrix $A$, define $A_B=Q^+AQ$ and $m_A=A_B^T A_B$. The Makhlin invariants are:

\begin{equation}
G_1=\frac{\Tr^2(m_A)}{16\det A},
\qquad
G_2=\frac{\Tr^2(m_A)-\Tr(m_A^2)}{4\det A}.
\label{eq:g1g2}
\end{equation}

They do not change under local wrappers; for $[\CNOT]$ we get $(G_1,G_2)=(0,1)$. The numerical measure of deviation from the target class $[\CNOT]$ is:

\begin{equation}
d_{[\CNOT]}=|G_1|^2+|G_2-1|^2.
\label{eq:d_def}
\end{equation}

\subsection{Fibonacci anyons}

In three-dimensional space, identical particles come in only two varieties: when two particles exchange, the wave function either remains unchanged (bosons) or changes sign (fermions). The reason is topological: in 3D, the trajectory of a double exchange is contractible to a point, so the square of the exchange operator must be the identity, and its eigenvalues must be $\pm1$.

In two-dimensional systems (electron gas in a strong magnetic field, films, heterostructures) this restriction is lifted: the world lines of particles in $(2+1)$-dimensional space-time form \emph{braids} that cannot be untangled by continuous deformation~\cite{LeinaasMyrheim1977,NayakRMP2008}. The exchange of two particles becomes a topologically nontrivial operation, and the wave function can acquire an arbitrary phase $e^{i\theta}$ --- hence the name \emph{anyons} (from "any"). Moreover, if the ground state of a system with fixed particle positions is degenerate (has dimension greater than~$1$), the exchange acts on the degeneracy space as an entire unitary matrix, not as a phase. Successive exchanges are multiplied like matrices and, generally speaking, do not commute. Such anyons are called non-Abelian, and they are suitable for computation: the logical operation corresponds to the physical braiding of quasiparticles around each other, and the result depends only on the topology of the world line braid—local trajectory deformations, jitter, and control inaccuracies have no effect. This is topological protection at the level of the gate itself.

The second building block of anyon theory is fusion: two anyons can be brought close together and treated as a single composite particle. Rules of the form $a\times b=\sum_c N_{ab}^{c}\,c$ list the possible outcomes. For non-Abelian anyons, there are multiple outcomes, and the state vector in the space of possible fusion outcomes is the carrier of quantum information: it is nonlocal (distributed between the anyons) and inaccessible to local measurements.

The Fibonacci model is the simplest universal non-Abelian anyon model~\cite{RowellStongWang2009}. It has a single non-trivial anyon $\tau$ (and trivial vacuum $\mathbf1$) and a single fusion rule:

\begin{equation}
\tau\times\tau=\mathbf1+\tau .
\label{eq:fusion}
\end{equation}

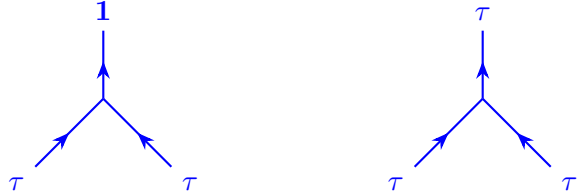
\begin{figure}[h]
\centering
\begin{tikzpicture}[thick,scale=0.9]
  \draw[-Stealth, blue, thick] (-1,-1) node[below left]{$\tau$} -- (-0.5,-0.5);
  \draw[blue, thick] (0,0) -- (-0.6,-0.6);
  \draw[-Stealth, blue, thick] (1,-1) node[below right]{$\tau$} -- (0.5,-0.5);
  \draw[blue, thick] (0,0) -- (0.6,-0.6);
  \draw[-Stealth, blue, thick] (0,0) -- (0,0.55);
  \draw[blue, thick] (0,0.45) -- (0,1) node[above]{$\mathbf1$};
\end{tikzpicture}
\hspace{2cm}
\begin{tikzpicture}[thick,scale=0.9]
  \draw[-Stealth, blue, thick] (-1,-1) node[below left]{$\tau$} -- (-0.5,-0.5);
  \draw[blue, thick] (0,0) -- (-0.6,-0.6);
  \draw[-Stealth, blue, thick] (1,-1) node[below right]{$\tau$} -- (0.5,-0.5);
  \draw[blue, thick] (0,0) -- (0.6,-0.6);
  \draw[-Stealth, blue, thick] (0,0) -- (0,0.55);
  \draw[blue, thick] (0,0.45) -- (0,1) node[above]{$\tau$};
\end{tikzpicture}
\caption{Two fusion channels of a Fibonacci anyon pair, $\tau\times\tau=\mathbf1+\tau$: vacuum (left) and $\tau$ (right). The state vector in outcome space is the carrier of quantum information.}
\label{fig:bg-fusion}
\end{figure}

Two $\tau$-anyons can merge either into a vacuum or back into $\tau$ --- a two-dimensional space of outcomes Fig.~\ref{fig:bg-fusion}.

In Appendix~\ref{app:fibonacci} we show that the dimension of the fusion space of $n$ anyons $\tau$ with total charge $\mathbf{1}$ grows like the Fibonacci numbers: $1,1,2,3,5,8,\dots$ and, in general, has an asymptotic form of dimension $\dim V_{\tau^n}^{\mathbf1}=\Theta(\varphi^n)$, so $\varphi$ is called the quantum dimension of the anyon, which is equal to the golden ratio. The golden ratio permeates the entire arithmetic of the model -- in particular, it defines the ring of quantized leakage values in lemma~\ref{lem:quant}. The key property of the model is braid universality: any unitary operator can be approximated by braids alone (without additional, topologically unprotected operations). A possible physical realization of the Fibonacci topological order is considered to be the fractional quantum Hall effect state at $\nu=12/5$, associated with the Read--Rezayi phase at $k=3$; however, its experimental identification is not yet definitive \cite{ReadRezayi1999,Mong2017}. Regardless of this, the doubled Fibonacci topological order admits an exact theoretical realization in the Levin--Wen lattice string-net models \cite{LevinWen2005}.

\subsubsection{Artin relations}

The braiding process of $n$ anyons arranged in a row is described by the braid group $B_n$. Its generators $\sigma_1,\dots,\sigma_{n-1}$ are elementary exchanges of adjacent anyons ($\sigma_i$ exchanges anyons $i$ and $i{+}1$ clockwise), and the full set of defining relations was established by Artin~\cite{Artin1947}:

\begin{align}
\sigma_i\,\sigma_j&=\sigma_j\,\sigma_i, && |i-j|\ge2 \quad\text{(distant exchanges are independent)},
\label{eq:artin-far}\\
\sigma_i\,\sigma_{i+1}\,\sigma_i&=\sigma_{i+1}\,\sigma_i\,\sigma_{i+1}, &&
\text{(braid relation)}.
\label{eq:artin-braid}
\end{align}

The second relation is precisely the Yang-Baxter equation~\cite{NayakRMP2008}. Geometrically, both mean that the braids that can be transformed into each other by a continuous deformation of the threads define the same element of the group. The anyon model associates with each generator a unitary matrix acting on the fusion space -- the representation $\rho:B_n\to U(d)$~\cite{Kitaev2003,NayakRMP2008,HadjiivanovGeorgiev2024}; the word $w=\sigma_{i_1}\dots\sigma_{i_L}$ of length $L$ goes into the product of matrices $\rho(w)$.

For algebraic constructions of braid-group representations using $R$-matrices and Hecke algebras, see~\cite{Jones1987Hecke,AnokhinaMorozov2014Cabling}, in particular Section~3.3 of the latter.

\subsubsection{\texorpdfstring{$F$- and $R$-matrices}{F- and R-matrices} \label{sec:F_and_R}}

To write $\rho(\sigma_i)$ explicitly, we need a basis for the fusion space. It is constructed using \emph{fusion trees}: the order in which anyons merge pairwise is fixed and intermediate outcomes are labeled. Two elementary objects completely define the model:

\begin{itemize}
\item \emph{$R$-matrix} --- the phase of exchange of two adjacent anyons with a known outcome of their fusion. For Fibonacci
\begin{equation}
R=\begin{pmatrix} e^{-4\pi i/5}&0\\ 0&e^{3\pi i/5}\end{pmatrix}
\label{eq:R}
\end{equation}
in the basis of outcomes $\{\mathbf1,\tau\}$: a pair merging into a vacuum acquires phase $e^{-4\pi i/5}$ upon exchange, while merging into $\tau$ acquires phase $e^{3\pi i/5}$.
\item \emph{$F$-matrix} --- basis change between two orders of fusion of three anyons, $(a\,b)\,c\leftrightarrow a\,(b\,c)$ (associator). For Fibonacci
\begin{equation}
F=\begin{pmatrix} \varphi^{-1}&\varphi^{-1/2}\\[2pt] \varphi^{-1/2}&-\varphi^{-1}\end{pmatrix},
\qquad F^2=I , \qquad \text{where} \qquad \varphi = \frac{1+\sqrt{5}}{2}
\label{eq:F}
\end{equation}
\end{itemize}

The consistency of $F$ and $R$ with each other (the equations of the pentagon and hexagon) fixes the model uniquely up to complex conjugation. The exchange of \emph{non-adjacent} anyons along the tree is written as brackets $FRF$: go to the basis where the pair is adjacent, exchange, return.

\subsection{The \texorpdfstring{$\ket{\mathrm{NC}}$}{|NC>} state and leakage}\label{sec:bg-encoding}

One logical qubit is conventionally encoded in three anyons $\tau$ with a two-dimensional fusion space. For two qubits, $6$ anyons are taken with a total charge $\mathbf {1}$. Here, a circumstance central to our work arises: the fusion space of $\tau^6$ has dimension $\mathrm{Fib}(6-1)=5$ see Appendix~\ref{app:fibonacci}, whereas two qubits require only $4=2\times2$ states. The fifth non-computational state $\NC$ is irremovable. In the \emph{caterpillar} basis of fusion trees, following~\cite{CuiWang2019} (the same encoding is used
in~\cite{Burke2024}; explicit $F,R$-matrices can be found in \cite{HadjiivanovGeorgiev2024}), we choose the order of bases as follows:

\begin{equation}
\{\;\NC,\;\ket{11},\;\ket{1\tau},\;\ket{\tau1},\;\ket{\tau\tau}\;\},
\label{eq:fib_basis}
\end{equation}

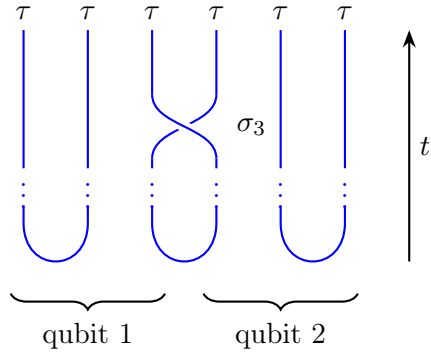
\begin{figure}[h]
\centering
\begin{tikzpicture}[thick,scale=0.85]
  \newanyon(0,0) \newanyon(2,0) \newanyon(4,0)
  \foreach \x in {0,1,4,5} {
    \draw[blue,thick] (\x,0) -- (\x,0.25);
    \draw[blue,thick] (\x,0.85) -- (\x,3.0);
  }
  \foreach \x in {2,3} {
    \draw[blue,thick] (\x,0) -- (\x,0.25);
    \draw[blue,thick] (\x,0.85) -- (\x,1.0);
  }
  {%
    \foreach \x in {0,...,5}
      \node[blue,inner sep=0pt] at (\x,0.55) {$\vdots$};
    \draw[-Stealth,black,thick] (6.0,-0.6) -- (6.0,3.0)
      node[midway,right,text=black] {$t$};
  }
  \rcross(2,1.0)
  \node[right] at (3.15,1.5) {$\sigma_3$};
  \draw[blue,thick] (2,2.0) -- (2,3.0);
  \draw[blue,thick] (3,2.0) -- (3,3.0);
  \foreach \x in {0,...,5} \node[above] at (\x,3.0) {$\tau$};
  \draw[decorate,decoration={brace,mirror,amplitude=6pt}] (-0.2,-1.1) -- (2.2,-1.1)
       node[midway,below=7pt] {qubit 1};
  \draw[decorate,decoration={brace,mirror,amplitude=6pt}] (2.8,-1.1) -- (5.2,-1.1)
       node[midway,below=7pt] {qubit 2};
\end{tikzpicture}
\caption{A two-qubit register of $6$ $\tau$ anyons (produced by three pairs from the vacuum; time flows upwards) and the action of the generator $\sigma_3$ --- the exchange of anyons $3$ and $4$. $\sigma_3$ is the only generator crossing the boundary of qubits: only it entangles them and only it hooks the non-computational state $\NC$, Eq.~\eqref{eq:sigma3}.}
\label{fig:bg-register}
\end{figure}

The labels inside the kets are the fusion charges of the pairs $(1,2)$ and $(5,6)$ of anyons (the subscript $0$ everywhere below corresponds to $\NC$). The five generators $\sigma_1,\dots, \sigma_5\in\Bsix$ become explicit unitary matrices $5\times5$ of the following structure:

\begin{itemize}
\item $\sigma_1,\sigma_5$ --- \emph{diagonal}: exchange within a pair whose charge is already a base label; the diagonal contains phases from Eq.~\ref{eq:R}. The spectrum of each is $\{e^{3\pi i/5}\ (\times3),\ e^{-4\pi i/5}\ (\times2)\}$.

\begin{equation}
\scalebox{0.86}{$\begin{gathered}
\sigma_1=
\begin{pmatrix}
e^{3\pi i/5}&0&0&0&0\\
0&e^{-4\pi i/5}&0&0&0\\
0&0&e^{-4\pi i/5}&0&0\\
0&0&0&e^{3\pi i/5}&0\\
0&0&0&0&e^{3\pi i/5}
\end{pmatrix},
\sigma_5=
\begin{pmatrix}
e^{3\pi i/5}&0&0&0&0\\
0&e^{-4\pi i/5}&0&0&0\\
0&0&e^{3\pi i/5}&0&0\\
0&0&0&e^{-4\pi i/5}&0\\
0&0&0&0&e^{3\pi i/5}
\end{pmatrix}
\end{gathered}
$}
\label{eq:sigma15}
\end{equation}

\item $\sigma_2,\sigma_4$ --- \emph{block} $2\times2$ brackets $FRF$ acting on single-qubit pairs of labels ($\sigma_2$ mixes $\ket{11}\! \leftrightarrow\!\ket{\tau1}$ and $\ket{1\tau}\!\leftrightarrow\!\ket{\tau\tau}$; $\sigma_4$ --- similarly in the second qubit). They \emph{do not} entangle the qubits and do not affect the $\NC$ state.

\begin{equation}
\scalebox{0.86}{$\begin{gathered}
\sigma_2=
\begin{pmatrix}
e^{3\pi i/5}&0&0&0&0\\
0&\varphi^{-1}e^{4\pi i/5}&0&\varphi^{-1/2}e^{-3\pi i/5}&0\\
0&0&\varphi^{-1}e^{4\pi i/5}&0&\varphi^{-1/2}e^{-3\pi i/5}\\
0&\varphi^{-1/2}e^{-3\pi i/5}&0&-\varphi^{-1}&0\\
0&0&\varphi^{-1/2}e^{-3\pi i/5}&0&-\varphi^{-1}
\end{pmatrix}, \\
\sigma_4=
\begin{pmatrix}
e^{3\pi i/5}&0&0&0&0\\
0&\varphi^{-1}e^{4\pi i/5}&\varphi^{-1/2}e^{-3\pi i/5}&0&0\\
0&\varphi^{-1/2}e^{-3\pi i/5}&-\varphi^{-1}&0&0\\
0&0&0&\varphi^{-1}e^{4\pi i/5}&\varphi^{-1/2}e^{-3\pi i/5}\\
0&0&0&\varphi^{-1/2}e^{-3\pi i/5}&-\varphi^{-1}
\end{pmatrix}.
\end{gathered}
$}
\label{eq:sigma24}
\end{equation}

\item $\sigma_3$ --- \emph{the only entangling} generator and simultaneously \emph{the only} one that has matrix elements between the code and $\NC$: it acts as an $FRF$-block on the pair $\{\NC,\ket{\tau\tau}\}$,

\begin{equation}
\scalebox{0.86}{$\begin{gathered}
\sigma_3=
\begin{pmatrix}
\varphi^{-1}e^{4\pi i/5}&0&0&0&\varphi^{-1/2}e^{-3\pi i/5}\\
0&e^{-4\pi i/5}&0&0&0\\
0&0&e^{3\pi i/5}&0&0\\
0&0&0&e^{3\pi i/5}&0\\
\varphi^{-1/2}e^{-3\pi i/5}&0&0&0&-\varphi^{-1}
\end{pmatrix},
\end{gathered}
$}
\label{eq:sigma3}
\end{equation}

and phases in the remaining three states.
\end{itemize}
This structure illustrates the reason for the special attention to leakage: entanglement and leakage enter the model via the same matrix element of magnitude $\varphi^{-1/2}$ in Eq.~\ref{eq:sigma3}. Without $\sigma_3$, the qubits cannot be entangled; using it inevitably entangles $\NC$, Fig.~\ref{fig:bg-register}.

\begin{definition}[Leakage]\label{def:leakage}

Let $U=\rho(w)$ be the braid matrix, and $A=U_{[1:,1:]}$ be its $4\times4$ computational unit. If the computational state $\ket j$ is fed to the input, the probability of being in $\NC$ at the output is $|U_{0j}|^2$. Unitarity of the zero-th row, $|U_{00}|^2+\sum_{j\ge1}|U_{0j}|^2=1$, collapses the total leakage into a single matrix element:

\begin{equation}
\ell(U)\;=\;1-|U_{00}|^2\;=\;\sum_{j\ge1}|U_{0j}|^2 .
\label{eq:bg-leak}
\end{equation}
\end{definition}

For $|U_{00}|=1$, the matrix is block-diagonal ($\mathbf1\oplus4$): the non-computational state is isolated, and only the unobservable phase remains on it. Crucially, leakage is not a typical gate error: it takes the state out of the code space, and standard error-correcting codes do not correct it~\cite{Suchara2015,AliferisTerhal2007}. Whether a finite braid word in the standard six-Fibonacci-anyon encoding can implement an exactly leakage-free entangling gate using braiding alone remains an open question~\cite{AinsworthSlingerland2011,BigelowLevaillant2019,CuiWang2019,Burke2024}, the practical problem is therefore to minimize leakage at fixed braid length.

Thus, the two-qubit compilation problem reduces to finding a positive word of fixed length whose computational unit approximates the local class $[\CNOT]$, while the leakage remains minimal. The exact search space, the generator matrices used, and the optimization criteria are formalized in Section ~\ref{sec:setup}.

\section{Main experiment pipeline}\label{sec:setup}

\subsection{Problem statement}

Since we have already defined the generators of the groups Eqs.\eqref{eq:sigma15},\eqref{eq:sigma24},\eqref{eq:sigma3}, it remains to obtain a lower bound for the attainable leakage. First, we show that, for a fixed length, a nonzero leakage cannot arbitrarily tend to zero, since it lies in $\mathbb Z[\varphi]$, the ring of integers of $\mathbb Q(\sqrt5)$.

\begin{lemma}[Leakage quantization]\label{lem:quant}
Let $U$ be the matrix of a positive word in $\{\sigma_1,\sigma_2,\sigma_3,\sigma_4,\sigma_5\}$, and let $\zeta_5=e^{2\pi i/5}$. Then
\begin{equation}
U_{00}\in\mathbb Z[\zeta_5],
\qquad
\ell(U)=1-|U_{00}|^2\in\mathbb Z[\varphi].
\end{equation}
\end{lemma}

\begin{proof}
Since $\varphi^{-1}=\varphi-1$ and
$\varphi^{-1/2}=\sqrt\varphi\,(\varphi-1)$, all generator entries
belong to $\mathbb Z[\zeta_5,\sqrt\varphi]$.

In the chosen basis, let $\gamma=(1,0,1,1,0)$. Direct inspection of
the generators shows that their $(k,l)$ entries belong to
$\mathbb Z[\zeta_5]$ when $\gamma(k)+\gamma(l)$ is even, and to
$\sqrt\varphi\,\mathbb Z[\zeta_5]$ when it is odd.
This property is preserved under matrix multiplication: in every
summand $A_{km}B_{ml}$, the parity is
\begin{equation}
\gamma(k)+2\gamma(m)+\gamma(l)
\equiv\gamma(k)+\gamma(l)\pmod2,
\end{equation}
and $(\sqrt\varphi)^2=\varphi\in\mathbb Z[\zeta_5]$.
Thus $U_{00}\in\mathbb Z[\zeta_5]$, because $2\gamma(0) \equiv 0$, and consequently
\begin{equation}
|U_{00}|^2\in
\mathbb Z[\zeta_5]\cap\mathbb R
=\mathbb Z[\varphi].
\end{equation}
It follows that $1-|U_{00}|^2=\ell(U)\in\mathbb Z[\varphi]$.
\end{proof}

\begin{lemma}[Lower bound on nonzero leakage]
\label{lem:nonzero-leakage-bound}
Let $U$ be the matrix of a positive word of length $L$ in
$\{\sigma_1,\sigma_2,\sigma_3,\sigma_4,\sigma_5\}$.
Let $g:\mathbb Q(\zeta_5,\sqrt\varphi)\to\mathbb C$ be the field
embedding determined by
\begin{equation}
g(\zeta_5)=\zeta_5^2,
\qquad
g(\sqrt\varphi)=\frac{i}{\sqrt\varphi}.
\end{equation}
Its restriction to $\mathbb Q(\sqrt5)$ sends $\sqrt5$ to $-\sqrt5$ -- standard Galois conjugation.
Set
\begin{equation}
C=\max_{1\leq j\leq5}\|\sigma_j^g\|_2=2.8901\ldots,
\end{equation}
where $g$ is applied entrywise and $\|\cdot\|_2$ is the spectral
operator norm. If $\ell(U)>0$, then
\begin{equation}
\ell(U)\geq\frac{1}{1+C^{2L}}
\geq\frac12 C^{-2L}
\end{equation}
\end{lemma}

\begin{proof}
By Lemma~\ref{lem:quant}, we can write
\begin{equation}
\ell(U)=\frac{a+b\sqrt5}{2},
\qquad
a,b\in\mathbb Z,\quad a-b\equiv0\pmod2
\end{equation}
If $\ell(U)>0$, its algebraic norm is a nonzero integer:
\begin{equation}
\ell(U)|g(\ell(U))| = |\ell(U)\,g(\ell(U))|
=\left|\frac{a^2-5b^2}{4}\right|\in\mathbb N
\end{equation}
Therefore,
\begin{equation}
\ell(U)\geq\frac{1}{|g(\ell(U))|}
\end{equation}

On $\mathbb Q(\zeta_5)$, the map $g$ commutes with complex
conjugation because the Galois group is Abelian. Since
$U_{00}\in\mathbb Z[\zeta_5]$, this gives
\begin{equation}
g(\ell(U))=1-|g(U_{00})|^2.
\end{equation}
Moreover, applying $g$ entrywise preserves matrix products, so
\begin{equation}
|g(U_{00})|
=|(U^g)_{00}|
\leq\|U^g\|_2
=\left\|\prod_{s=1}^{L}\sigma_{i_s}^g\right\|_2
\leq\prod_{s=1}^{L}\|\sigma_{i_s}^g\|_2
\leq C^L.
\end{equation}
Combining these estimates yields
\begin{equation}
\ell(U)
\geq
\frac{1}{\left|1-|g(U_{00})|^2\right|}
\geq
\frac{1}{1+|g(U_{00})|^2}
\geq
\frac{1}{1+C^{2L}}
\geq
\frac12 C^{-2L}
\end{equation}
where the last inequality uses $C\geq1$.
\end{proof}

\subsubsection{Identity of our leakage and the Burke et al. measure}

In this paper, leakage is defined as
\begin{equation}
\ell(U)=1-|U_{00}|^2,
\end{equation}
while Burke et al.~\cite{Burke2024} use the trace norm
\begin{equation}
d_U(A)=\|A^\dagger A-I\|_1
\end{equation}
for computational block $A$. For the block structure under consideration, the matrix $A^+ A-I$ has rank at most one, and its only nonzero singular value is $1-|U_{00}|^2$. Therefore,
\begin{equation}
\ell(U)=d_U(A).
\end{equation}
Thus, our values of $\ell(U)$ are directly comparable with the values of $d_U(A)$ from~\cite{Burke2024}.

\subsection{Naive exhaustive search}

The direct algorithm enumerates all $q^L$ words of the alphabet of cardinality $q$, sequentially constructs the full matrix $U$ for each, and computes the leakage and Makhlin invariants. For the base alphabet $q=5$, this requires $\Theta(L5^L)$ small matrix multiplications. The search is chunked and therefore does not exhaust GPU memory, but its time grows exponentially with $L$, of order $\Theta(L5^L)$, taking into account the multiplication of all matrices in the braid. This implementation serves as a check baseline for MITM on accessible lengths.

\subsection{Distribution function of operators \texorpdfstring{$F_L(\ell)$}{F\_L(ell)} over leakage}

Before the main experiment, we will conduct several experiments to study the distribution function $F_L(\ell)$ of operators generated by braid length $L$ from $3$ generators $\{\sigma_1,\sigma_2,\sigma_3\}$ or $5$ generators $\{\sigma_1,\sigma_2,\sigma_3,\sigma_4,\sigma_5\}$, respectively.

We define the cumulative distribution function (CDF) of leakage as follows:

\begin{definition}[Cumulative distribution function (CDF) of leakage $F$]\label{def:CDF}
Let $\mathcal A_q$ be an alphabet from $q$ generators, and let the word $w$ be chosen uniformly from all words of length $L$ over this alphabet. Then we define the leakage distribution function as

\begin{equation}
F_{L}^{\mathcal A_q}(\ell_0)
=
\Pr\!\left(
\ell\!\left(\rho(w)\right)<\ell_0
\right)
=
\frac{
\left|
\left\{
w\in\mathcal A_q^L:
\ell\!\left(\rho(w)\right)<\ell_0
\right\}
\right|
}{
\left|\mathcal A_q^L\right|
}
=
\frac{
\left|
\left\{
w\in\mathcal A_q^L:
\ell\!\left(\rho(w)\right)<\ell_0
\right\}
\right|
}{
q^L
}.
\label{eq:main-leak-cdf}
\end{equation}
Different words that define the same matrix are counted separately.

\end{definition}

This will help us understand the difficulty of the problem in the general case: how rare are low-leakage operators generated by such products and how the difficulty of finding them grows with increasing $L$. To this end, for each $L$ presented in the graphs Fig.~\ref{fig:structured-leakage-cdf}, we will generate all possible braids generated by these generators: $3^L$ and $5^L$ for the left and right graphs, respectively. Each curve represents the fraction of words of length $L$ whose leakage is below a given value $\ell$. Since $\sigma_3$ is the only generator that directly couples the non-computational state to the computational sector, every word that does not contain $\sigma_3$ has zero leakage. This accounts for the pronounced point mass at zero for small $L$. Each finite-length CDF is monotone in $\ell$ by definition.

For the three-generator alphabet $\{\sigma_1,\sigma_2,\sigma_3\}$ (left panel of Fig.~\ref{fig:structured-leakage-cdf}), the finite-length CDFs cross the limiting quadratic curve and approach it from both sides. Thus, there is no fixed pointwise ordering between the finite-length curves and their limit. Fitting the $L=29$ curve by a power law $\ell^\alpha$ gives $\alpha\approx2.004$, in close agreement with the limiting law proved in Appendix~\ref{app:leak_CDF}:

\begin{equation}
F_{3,\infty}(\ell)=\ell^2.
\label{eq:3gen_F_ing}
\end{equation}

For the full five-generator alphabet $\{\sigma_1,\sigma_2,\sigma_3,\sigma_4,\sigma_5\}$ (right panel of Fig.~\ref{fig:structured-leakage-cdf}), fewer lengths are computationally accessible and the convergence is therefore less developed. Over the observed range, the limiting quartic curve lies below the finite-length CDFs, which approach it from above as $L$ increases. This behavior is consistent with the second limiting law proved in Appendix~\ref{app:leak_CDF}:

\begin{equation}
F_{5,\infty}(\ell)=\ell^4.
\label{eq:5gen_F_ing}
\end{equation}

\def\showRestricted{2,6,10,14,18,21,23,25,27,29}
\def\showFullFive{2,4,6,8,10,12,14,16,18,20}

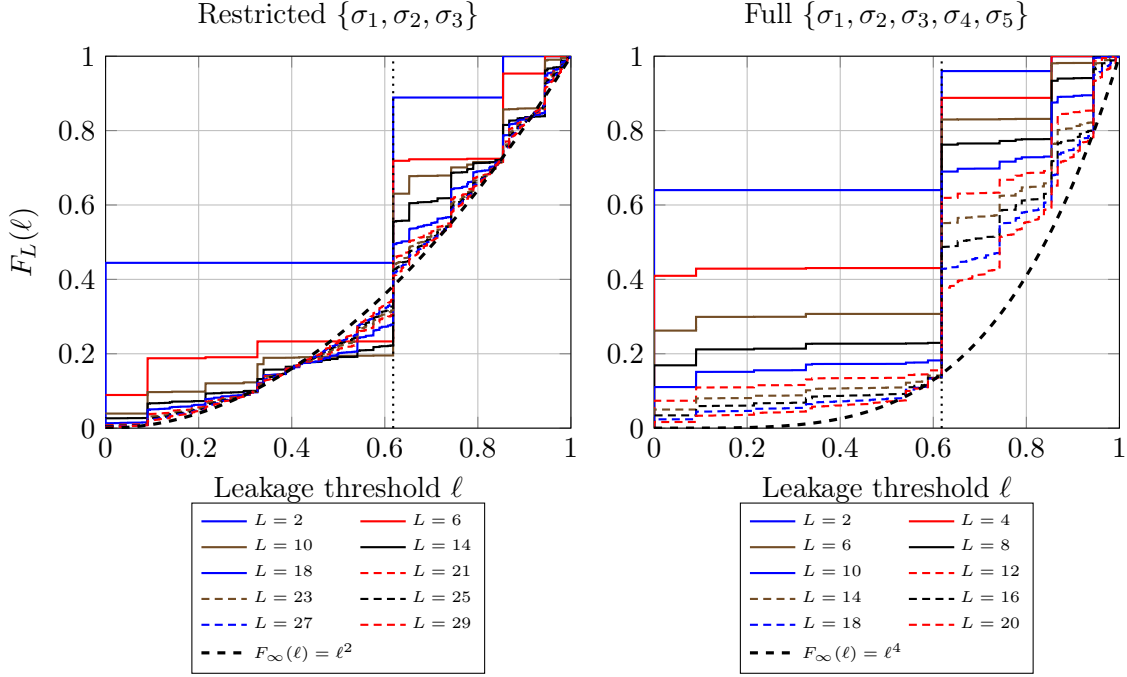
\begin{figure}[ht!]
    \centering
    \begin{tikzpicture}
    \begin{groupplot}[
      group style={
        group size=2 by 1,
        horizontal sep=1.1cm
      },
      width=0.455\textwidth,
      height=6.5cm,
      xmin=0,
      xmax=1,
      ymin=0,
      ymax=1,
      xlabel={Leakage threshold $\ell$},
      grid=both,
      no marks,
      every axis plot/.append style={thick},
      legend cell align=left,
      legend style={
        font=\tiny,
        legend columns=2,
        at={(0.5,-0.20)},
        anchor=north
      }
    ]
    \nextgroupplot[
      title={Restricted $\{\sigma_1,\sigma_2,\sigma_3\}$},
      ylabel={$F_L(\ell)$}
    ]
    \foreach \L in \showRestricted {
      \addplot+
        table[
          x=tau,
          y=fraction_below_tau,
          col sep=comma
        ]{cdf/cdf_restricted_L\L.csv};
    \addlegendentryexpanded{$L=\L$}
    }
    \addplot[
      black,
      dashed,
      very thick,
      domain=0:1,
      samples=100
    ]{x^2};
    \addlegendentry{$F_\infty(\ell)=\ell^2$}

    \draw[
      black,
      dotted,
      thick
    ]
      (axis cs:0.6180339887498949,0)
      --
      (axis cs:0.6180339887498949,1);

    \nextgroupplot[
      title={Full $\{\sigma_1,\sigma_2,\sigma_3,\sigma_4,\sigma_5\}$}
    ]

    \foreach \L in \showFullFive {
      \addplot+
        table[
          x=tau,
          y=fraction_below_tau,
          col sep=comma
        ]{cdf/cdf_full5_L\L.csv};
      \addlegendentryexpanded{$L=\L$}
    }

    \addplot[
      black,
      dashed,
      very thick,
      domain=0:1,
      samples=100
    ]{x^4};
    \addlegendentry{$F_\infty(\ell)=\ell^4$}

    \draw[
      black,
      dotted,
      thick
    ]
      (axis cs:0.6180339887498949,0)
      --
      (axis cs:0.6180339887498949,1);

    \end{groupplot}
    \end{tikzpicture}

    \caption{Exhaustive leakage CDFs for selected lengths in the positive
    restricted alphabet $\{\sigma_1,\sigma_2,\sigma_3\}$ (left) and the positive
    full alphabet $\{\sigma_1,\ldots,\sigma_5\}$ (right). Every word of each
    displayed length is counted with multiplicity. As the length increases, the
    finite discrete distributions approach the respective Haar limits
    $F_\infty(\ell)=\ell^2$ and $F_\infty(\ell)=\ell^4$ proved in
    Appendix~\ref{app:leak_CDF}. The black dotted lines mark
    $\ell=\varphi^{-1}$. The finite-length curves were obtained by exhaustive
    enumeration in production machine arithmetic; they illustrate convergence
    but are not used to prove Haar mixing.}
    \label{fig:structured-leakage-cdf}
\end{figure}

Thus, a numerical experiment leads to a hypothesis, but does not prove it. We have included its mathematical proof in Appendix~\ref{app:leak_CDF}. Here we will simply present some considerations of this proof.

The main idea of the proof consists of two steps. First, after factoring the physically irrelevant global phase, the distribution of long positive-only random words converges to the Haar measure on the corresponding projective unitary group $U(d)/U(1)$ acting on the invariant subspace of the set of generators. This is achieved using the density of the projective image of the Jones representation~\cite{FreedmanLarsenWang2002Density,Kuperberg2011} and the Kawada--Ito theorem together with its modern formulations \cite{KawadaIto1940,Mustafayev2022,Monakov2025}. Quantitative questions on the rate of such convergence for random walks on compact groups have been studied, in particular, in~\cite{Bhattacharya1972}. However, the rate of convergence for the ensembles under consideration is not estimated there.

Second, for a Haar random operator in $d$ dimensions, the quantity:

\begin{equation}
X=|\langle\mathrm{NC}|U|\mathrm{NC}\rangle|^2
\end{equation}

has density $f_X(x)=(d-1)(1-x)^{d-2}$ for $0\leq x\leq1$. Therefore,

\begin{equation}
F_{\infty}(\ell) = \Pr\!\left(
1-X<\ell
\right)
=\ell^{d-1}.
\end{equation}

Hence,

\begin{equation}
\lim_{L\to\infty}F_L(\ell)=
\begin{cases}
\ell^2, & d=3,\\
\ell^4, & d=5.
\end{cases}
\end{equation}

Where $d$ is the dimension of the invariant subspace under the operator's actions. It is equal not to the number of generators, but to the dimension of the invariant subspace of generators.

\subsection{MITM-GEMM factorization}
\label{sec:mitm_formulation}

The basic computational idea is to split a word into two parts. For even length $L=2K$, we write:

\begin{equation}
    w=w_1w_2,
    \qquad
    U(w)=U_2U_1,
    \label{eq:w1w2}
\end{equation}

The generalization to odd $L$ is obvious: the left operator will be constructed from $K+1$ operators, and the right one from $K$ operators, where $L = 2K+1$. $U_1$ and $U_2$ correspond to half-braids of length $K$. For an alphabet of $q$ generators, the table of each half contains $q^K=q^{L/2}$ matrices. The construction of these tables and the storage of the necessary matrix data is henceforth called MITM-preprocessing.

The key advantage of splitting into two parts follows from the fact that the amplitude that determines the leakage is represented by a single bilinear convolution:

\begin{equation}
    (U_2U_1)_{00}
    =
    \sum_{i=0}^{4}(U_2)_{0i}(U_1)_{i0}.
    \label{eq:mitm_u00}
\end{equation}

Thus, for each pair, it is sufficient to multiply row $0$ of the left half-braid by column $0$ of the right half-braid, without computing the remaining $24$ elements of the full product matrix to find the leakage. For partitioning into more parts, this optimization fails, because all intermediate matrices must be computed in their entirety. For example, for partitioning into $3$ parts, all elements of the intermediate matrix, Eq.~\ref{eq:three_part_u00}, will be used.

\begin{equation}
    (U_3U_2U_1)_{00}
    =
    \sum_{i,j=0}^{4}
    (U_3)_{0i}(U_2)_{ij}(U_1)_{j0},
    \label{eq:three_part_u00}
\end{equation}

This two-half operation is performed by batch calls to GPU GEMM (\emph{GEneral Matrix-Multi\-plication}). The leakage of each pair is then $\ell=1-|(U_2U_1)_{00}|^2$.

Unlike classical collision-MITM~\cite{AmyMITM2013}, here factorization does not reduce the total number of braids checked, so the total search time remains asymptotic $\Theta(q^L)$, while preprocessing and memory require $\Theta(q^{L/2})$, so the gain is multiple constant rather than asymptotic.

The full matrices and Makhlin invariants are computed only for order $q^L F_L(\ell_L)$ of candidates that pass the adaptive filter $\ell<\ell_L$. This transfers the bulk of the exhaustive search to a cheap dense operation, and the expensive proximity test to $[\CNOT]$ is applied only to a small fraction of known low-leakage words.

\subsection{Searching Pipeline}

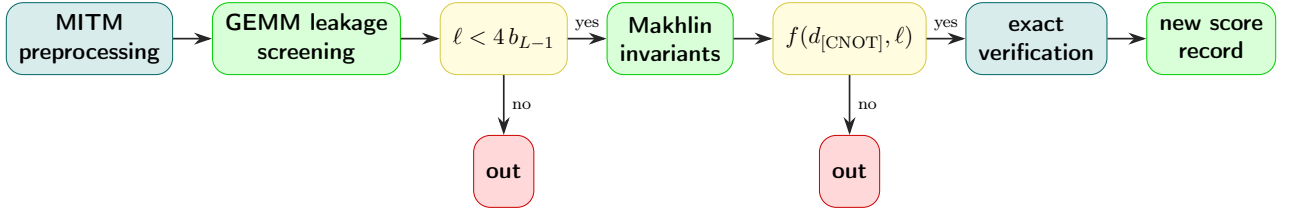
\begin{figure}[ht!!!]
\centering
\resizebox{\linewidth}{!}{
\begin{tikzpicture}[
  node distance=1.cm and .65cm,
  adaptive/.style={
    align=center,
    minimum width=0pt,
    inner xsep=6pt
  }
]

  \node (hmitm) [modern=teal, adaptive]
    {MITM\\preprocessing};

  \node (hgemm) [modern=green, adaptive, right=of hmitm]
    {GEMM leakage\\screening};

  \node (hleak) [modern=yellow, adaptive, right=of hgemm]
    {$\ell<4\,b_{L-1}$};

  \node (hmakhlin) [modern=green, adaptive, right=of hleak]
    {Makhlin\\invariants};

  \node (hscore) [modern=yellow, adaptive, right=of hmakhlin]
    {$f(d_{[\CNOT]},\ell)$};

  \node (hexact) [modern=teal, adaptive, right=of hscore]
    {exact\\verification};

  \node (hrecord) [modern=green, adaptive, right=of hexact]
    {new score\\record};

  \node (hout1) [modern=red, adaptive, below=of hleak]
    {out};

  \node (hout2) [modern=red, adaptive, below=of hscore]
    {out};

  \draw[arrowstyle] (hmitm) -- (hgemm);
  \draw[arrowstyle] (hgemm) -- (hleak);

  \draw[arrowstyle] (hleak) --
    node[above, font=\scriptsize] {yes} (hmakhlin);
  \draw[arrowstyle] (hleak) --
    node[right, font=\scriptsize] {no} (hout1);

  \draw[arrowstyle] (hmakhlin) -- (hscore);

  \draw[arrowstyle] (hscore) --
    node[above, font=\scriptsize] {yes} (hexact);
  \draw[arrowstyle] (hscore) --
    node[right, font=\scriptsize] {no} (hout2);

  \draw[arrowstyle] (hexact) -- (hrecord);

\end{tikzpicture}

}
\caption{A MITM-GEMM pipeline. The semibraids are enumerated independently, after which the values of $U_{00}(ab)=r_a\cdot c_b$ are batch-computed using GEMM. For candidates that pass the $\ell<4b_{L-1}$ filter, the Makhlin invariants are calculated and one of two optimization criteria is applied: $\min S=d+4\ell$ or $\min\ell$ for $d_{[\CNOT]}<10^{-4}$. The finalists undergo a rigorous check.}
\label{fig:mitm-gemm-pipelines}
\end{figure}

Throughout our experiments, we used two possible pipelines. The substantive results will indicate which method was used depending on the form of the function $f(d_{[\CNOT]},\ell)$. We intentionally include all results, as both approaches to constructing $f(d_{[\CNOT]},\ell)$ effectively reduce leakage, but they track the distance to $[\CNOT]$ differently. This doesn't change the qualitative estimates, but each point corresponding to sufficiently large $L$ represents hours of computation, and in some cases, tens of hours of computation. For this reason, we don't rerun both methods over the entire length of the study.

Since in Section~\ref{sec:mitm_formulation} we have already introduced MITM-preprocessing and defined the GEMM algorithm, and Makhlin invariants are given in the introduction, in the following Sections we will talk about the adaptive choice of the leakage threshold $\ell_L$, its asymptotics, and also about the function $f(d_{[\CNOT]},\ell)$.

\subsubsection{Adaptive \texorpdfstring{$\ell_L$}{ell\_L} \label{sec:ada_tau}}

Let $b_{L-1}>0$ be the leakage of a verified word of length $L-1$ satisfying $d_{[\CNOT]\ L-1}+4b_{L-1} = d_{[\CNOT]\ L}+4b_{L}$. Adding a local leakage-free generator $\sigma_1$ preserves the leakage and Makhlin invariants, so there is a feasible word on length $L$ with the same leakage and $\ell^\star_L\le b_{L-1}$. Therefore, any threshold $\ell_L>b_{L-1}$ preserves the global minimum of the constrained problem in the chosen alphabet without an additional boundary constraint. We use $\ell_L=4b_{L-1}$ with an additional numerical guard band in machine filtering and $d$ --- filtering, assuming that we are looking for sufficiently accurate $[\CNOT]$ candidates. The threshold does not reduce the dense scanning of all pairs, but significantly reduces the number of full reconstructions and calculations of Makhlin invariants.

For example, as will be seen from the numerical experiment later, the asymptotics of the optimal leakage depends on the braid length $L$ as $\ell_{\mathrm{fit}} = \Theta(10^{-L/6})$. In this case, the proportion of candidates for which the Makhlin invariants are calculated has an asymptotic behavior:

\begin{equation}
P_{\mathrm{Makh}} = \Theta\left(10^{-\frac{L(d-1)}{6}}\right)
\label{eq:Mach_dole}
\end{equation}

Then the final asymptotics of the number of matrices for which it is necessary to calculate the Makhlin invariants is presented:

\begin{equation}
\mathbb E n_{\mathrm{Makh}} = \Theta\left(q^L10^{-\frac{L(d-1)}{6}}\right) = \Theta\left(\left(\frac{q}{10^{\frac{d-1}{6}}}\right)^L\right)
\label{eq:Mach_res}
\end{equation}

Which, in fact, for $5$ and $3$ generators without special optimizations, respectively, has asymptotics of order $\Theta(1.077^L)$ and $\Theta(1.393^L)$. That is, in fact, the main complexity asymptotics falls on the MITM preprocessing and GEMM stages, and the Makhlin invariants themselves, being the most computationally intensive stage per calculation, have a much slower growth rate of the number of these calculations, which ceases to be an obstacle to increasing $L$ and concentrates the entire complexity of the algorithm in the first two stages.

\subsubsection{\texorpdfstring{$f(d_{[\CNOT]},\ell)$}{f(d\_[CNOT],ell)} function}

This function is needed to simultaneously select braids close enough to $[\CNOT]$, but also without weakening the leakage. This function can be implemented in two ways:

\begin{equation}
f(d_{[\CNOT]},\ell)=\begin{cases}
S = d_{[\CNOT]}+4\ell, & S<S_{\mathrm{best}},\\[2pt]
d<10^{-4},  & \ell<b_{L-1}.
\end{cases}
\label{eq:f}
\end{equation}

From now on we will call these variants $S$-target and $\ell$-target, respectively.

\subsubsection{Pipeline description}

Now that we have an understanding of each stage of the computational pipeline, we can describe it stepwise. First, we take all matrices generated by given conditions, for example, a set of generators, and select those whose leakage is no more than $4$ times the best leakage over the previous length. Generally speaking, we exploit the fact that leakage reduction is always monotonic: if we had low leakage at the previous level, then adding any generator except $\sigma_3$ guarantees that the leakage will not change and we will not leave the $[\CNOT]$ class by definition. The factor $4$ is used to exclude situations with roundoff errors, but it does not worsen the asymptotics, and the proportion of discarded matrices remains high; see Section~\ref{sec:ada_tau}. This extensive step, which involves all possible matrices, is computationally the cheapest: we calculate only $1$ element of each matrix instead of $25$, and obtain each matrix by a single multiplication of all matrices with all matrices, each of which is obtained by multiplying $q^{L/2}$ matrices by $L/2$. This does not change the asymptotics itself, but significantly reduces the computational constant.

Next, we work only with a relatively small number of low-leakage operators. For them, we compute the Makhlin invariants, Eq.~\ref{eq:g1g2}, and then compute the distance to the target class, Eq.~\ref{eq:d_def}. Thus, for each known low-leakage braid, we obtain a pair $(d_{[\CNOT]}, \ell)$, which we can rank using the function $f(\cdot,\cdot)$, Eq.~\ref{eq:f}, and select the best candidate from the set of $d_{[\CNOT]}$ and $\ell$. Among this already rather small number of candidates, we perform an exact verification in high-bit arithmetic and derive the best candidate for $\ell$. This candidate is the new score record.

\section{Results}\label{sec:results}

In this Section, we present and analyze the results obtained using our methodology outlined in Section~\ref{sec:setup}, some of which were already announced above. We also provide a list of the optimizations we applied, their motivations, and an estimate of their speedup.

\subsection{Naive brute force vs MITM: \texorpdfstring{$L=14$ vs $L=21$}{L=14 vs L=21}}

\begin{figure}[tbp]
\centering
\resizebox{\linewidth}{!}{%
\begin{tikzpicture}
\begin{semilogyaxis}[
  width=\textwidth, height=9cm,
  xlabel={Braid length $L$}, ylabel={Leakage $\ell$},
  grid=both, xmin=8, xmax=74, ymin=1e-12,
  legend cell align=left,
  legend columns=3,
  legend style={at={(0.5,-0.17)}, anchor=north, font=\small,
                /tikz/every even column/.append style={column sep=14pt}},
]
\addplot[blue,mark=*,thick] coordinates {
  (10,0.190) (12,0.0476) (14,0.0476) (16,0.0476) (18,0.0201) (20,0.0053) (21,0.00394)
};
\addlegendentry{full-generators}

\addplot[red,mark=triangle*,dashed,thick] coordinates {
  (12,0.0476) (24,0.0178) (27,0.0220) (30,0.00756) (39,0.00474)
};
\addlegendentry{Burke et al. 2024~\cite{Burke2024}}

\addplot[gray,dashed,thick,domain=10:15,samples=100]
  {1/(1 + 2.890053638264^(2*x))};
\addlegendentry{floor (Lemma~\ref{lem:nonzero-leakage-bound})}

\addplot[green!50!black,mark=pentagon*,thick] coordinates {
  (22,3.945720761e-3)
  (23,9.626697932e-4)
  (24,9.626697932e-4)
  (25,9.626697932e-4)
  (26,9.626697932e-4)
  (27,9.626697932e-4)
  (28,6.619392914e-4)
  (29,1.137938492e-4)
  (30,1.137938492e-4)
  (31,6.556947085e-5)
  (32,6.556947085e-5)
  (33,6.556947085e-5)
  (34,6.556947085e-5)
};
\addlegendentry{restricted exhaustive}

\addplot[orange,mark=o,thick,mark size=2pt] coordinates {
  (35,1.137938492e-4)
  (36,3.856806156e-5)
  (37,3.856806156e-5)
  (38,2.754430157e-6)
  (39,2.754430157e-6)
  (40,2.754430157e-6)
  (41,2.754430157e-6)
  (42,5.432863324e-7)
  (43,5.432863324e-7)
  (44,5.432863324e-7)
  (45,5.432863324e-7)
  (46,5.432863324e-7)
  (47,5.432863324e-7)
  (48,5.432863324e-7)
  (49,5.432863324e-7)
  (50,2.955551335e-7)
  (51,2.955551335e-7)
  (52,5.632461638e-8)
  (53,5.632461638e-8)
  (54,5.632461638e-8)
  (55,1.540365373e-8)
  (56,1.540365373e-8)
  (57,6.224841055e-9)
  (58,6.224841055e-9)
  (59,6.224841055e-9)
  (60,6.224841055e-9)
  (61,4.415741365e-9)
  (62,4.415741365e-9)
  (63,1.384233806e-9)
  (64,2.124348426e-10)
  (65,2.124348426e-10)
  (66,2.124348426e-10)
  (67,2.124348426e-10)
  (68,2.124348426e-10)
  (69,8.783959780e-11)
  (70,4.809459212e-11)
  (71,4.445544161e-11)
  (72,4.445544161e-11)
};
\addlegendentry{palindromic quotient }

\addplot[
  black,
  dashdotted,
  very thick,
  domain=10:72,
  samples=150
] {10*sqrt(5)*exp(-ln(10)*x/6)};

\addlegendentry{
  $\ell_{\mathrm{fit}} =\sqrt5\,10^{1-L/6}$
}

\draw[red,dotted,thick]
  (axis cs:14,1e-12) -- (axis cs:14,1);

\node[red,anchor=south,rotate=90]
  at (axis cs:14,2e-9)
  {\footnotesize exact Burke et al.};

\draw[red,dotted,thick]
  (axis cs:8,4.74e-3) -- (axis cs:74,4.74e-3);

\node[red,anchor=south east]
  at (axis cs:73,4.74e-3)
  {\footnotesize Burke et al. best: $4.74\cdot10^{-3}$};

\end{semilogyaxis}
\end{tikzpicture}
}
\caption{Leakage of the best $[\CNOT]$ class gate as a function of braid length. The straight-line fit is chosen as the simplest combination of rational numbers and the golden ratio $\varphi$. The fraction of variance explained by the $\ell_{\mathrm{fit}}$ function is $R^2 = 0.9851$.}
\label{fig:curve}
\end{figure}
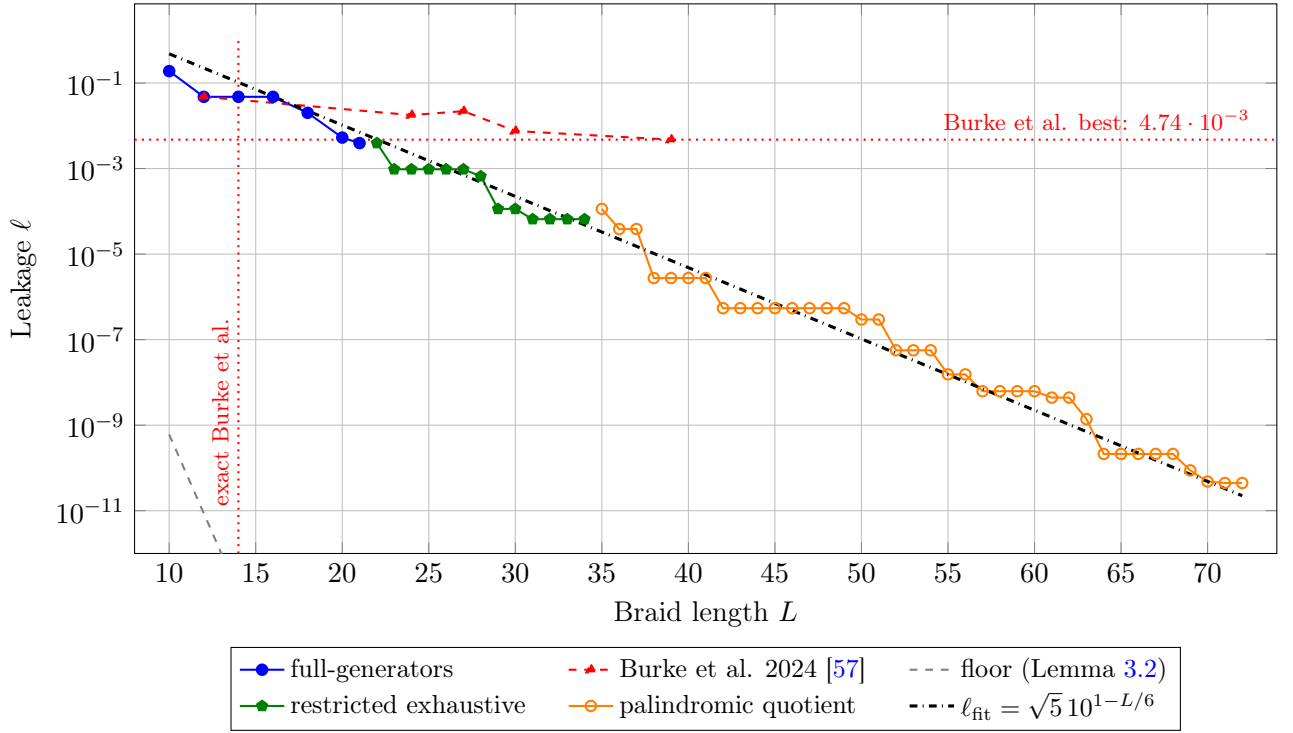

Tab.~\ref{tab:curve} presents a list of the most interesting points in the plot shown in Fig.~\ref{fig:curve}. Each Family in this table corresponds to a motivated narrowing of the search space. In the following Subsections, we introduce and explain the structure of each.

\begin{table}[H]
\centering
\scriptsize
\setlength{\tabcolsep}{2.5pt}
\renewcommand{\arraystretch}{1.12}

\begin{tabularx}{\textwidth}{
@{}
>{\centering\arraybackslash}p{0.40cm}
>{\raggedright\arraybackslash}p{1.30cm}
>{\centering\arraybackslash}p{1.10cm}
>{\centering\arraybackslash}p{1.60cm}
>{\centering\arraybackslash}p{1.60cm}
>{\ttfamily\tiny\raggedright\arraybackslash}X
>{\raggedright\arraybackslash}p{3.90cm}
@{}
}

\toprule
$L$
& Family
& $f$ target
& $\ell(U)$
& $d_{[\CNOT]}(U)$
& Representative word
& Comment
\\
\midrule

21
& full-5 MITM
& $S$
& $3.946{\cdot}10^{-3}$
& $7.779{\cdot}10^{-9}$
& 3 2 2 2 2 2 1 1 3 1 3 1 3 3 1 2 2 2 2 2 3
& Exhaustive over all $5^{21}$ positive words; historical composite-score optimum.
\\[2pt]

23
& restricted $\{1,2,3\}$
& $S$
& $9.627{\cdot}10^{-4}$
& $1.816{\cdot}10^{-5}$
& 3 2 2 2 2 1 1 1 1 1 2 2 1 3 3 3 3 2 2 2 2 2 3
& First strong record after generator reduction; exhaustive over $3^{23}$ words.
\\[2pt]

31
& boundary $3u3$
& $\ell$
& $6.557{\cdot}10^{-5}$
& $2.641{\cdot}10^{-5}$
& 3 3 2 2 2 2 1 3 2 2 2 3 3 1 3 3 1 1 1 1 1 1 2 2 3 3 2 2 2 3 3
& New leakage-primary record; exhaustive over all $3^{29}$ internal words.
\\[2pt]

38
& boundary palindrome
& $\ell$
& $2.754{\cdot}10^{-6}$
& $2.781{\cdot}10^{-8}$
& 3 3 3 3 3 2 2 1 1 2 2 3 3 3 2 2 3 1 2 2 1 3 2 2 3 3 3 2 2 1 1 2 2 3 3 3 3 3
& Literal palindrome $A\varepsilon A^R$; first large palindromic reduction; two exact-leak ties.
\\[2pt]

42
& boundary palindrome
& $\ell$
& $5.433{\cdot}10^{-7}$
& $4.104{\cdot}10^{-12}$
& 3 3 3 2 2 2 1 3 3 2 2 3 3 2 2 1 1 1 1 2 3 3 2 1 1 1 1 2 2 3 3 2 2 3 3 1 2 2 2 3 3 3
& Literal palindrome $A\varepsilon A^R$; nearly exact $[\CNOT]$ class; three exact-leak ties.
\\[2pt]

52
& $B_4^+$ quotient palindrome
& $\ell$
& $5.633{\cdot}10^{-8}$
& $1.635{\cdot}10^{-5}$
& 3 2 2 2 3 3 3 3 2 2 2 2 2 3 3 2 2 2 1 3 1 2 3 2 3 2 2 3 2 3 2 1 3 1 2 2 2 3 3 2 2 2 2 2 3 3 3 3 2 2 2 3
& Exact positive-$B_4$ quotient search; palindrome $A\varepsilon A^R$; leakage below $10^{-7}$.
\\[2pt]

55
& $B_4^+$ quotient palindrome
& $\ell$
& $1.540{\cdot}10^{-8}$
& $1.434{\cdot}10^{-5}$
& 3 2 2 2 1 3 3 3 2 2 2 2 3 3 2 2 2 2 2 2 1 1 2 2 1 1 1 1 1 1 1 2 2 1 1 2 2 2 2 2 2 3 3 2 2 2 2 3 3 3 1 2 2 2 3
& Exact quotient search with nonempty centre: $A\,1\,A^R$.
\\[2pt]

64
& $B_4^+$ quotient palindrome
& $\ell$
& $2.124{\cdot}10^{-10}$
& $1.884{\cdot}10^{-8}$
& 3 3 2 2 2 1 1 1 1 2 3 3 2 2 3 3 2 1 1 2 2 1 1 2 2 1 1 1 2 3 3 2 2 3 3 2 1 1 1 2 2 1 1 2 2 1 1 2 3 3 2 2 3 3 2 1 1 1 1 2 2 2 3 3
& Exact quotient palindrome $A\varepsilon A^R$; leakage reaches the $10^{-10}$ scale.
\\

\bottomrule
\end{tabularx}

\caption{Selected representatives from successive exhaustive search regimes.
Here $\ell(U)=1-|U_{00}|^2$, $d=d_{[\CNOT]}(U)$, and
$S=d+4\ell$. Leakage values are verified exactly, whereas $d$ is evaluated
numerically using Eq.~\ref{eq:d_def}. The historical searches at $L=21,23$
selected candidates by minimizing $S$; the searches represented by
$L\geq31$ minimize $\ell$ subject to $d<10^{-4}$. All listed representatives
satisfy this distance bound. The symbol $\varepsilon$ denotes an empty
central word.}

\label{tab:curve}

\end{table}

\subsection{Optimizations}\label{sec:scaling}

As Fig.~\ref{fig:curve} shows, the blue curve of the full five-generator search, directly comparable to the results of Burke et al.~\cite{Burke2024}, terminates at $L=21$. The dotted lines mark the boundary of their exhaustive search $L=14$ and the best sampling result at $L=39$. At $L=21$, our exhaustive search certifies the optimum of the historical composite score function $S=d+4\ell$; the found candidate has less leakiness than the best sampling result of Burke et al., despite the shorter braid length. Computing this point took about $68$ hours on $4\times$NVIDIA V$100$, so directly continuing the full five-generator search becomes impractical.

The further progress is based on $4$ successive steps: an empirically motivated reduction of the alphabet from $5$ to $3$ generators, fixing the boundary $\sigma_3$, moving to a palindromic family, and precisely removing duplicates of its halves via quotient over positive braid monoid relations.

\subsubsection{Generator reduction}

Based on the results of the exact five-generator search, it can be observed that for the run with the best leakiness, there are several equivalent braids; moreover, each corresponding leakiness level admits a representative written only in terms of $\{\sigma_1,\sigma_2,\sigma_3\}$. In particular, the champions $L=20,21$ already have this form. In this article, we will not delve into the mathematical side of this effect. We will only note that, from a physical perspective, such a set includes the diagonal operator $\sigma_1$, the entangling operator $\sigma_3$, and $\sigma_2$, which acts only on the first qubit. This may mean that all interactions with the second qubit are outside the entangling operators and are cancelled when moving on to consider the equivalence class $[\CNOT]$. Due to the empirically motivated alphabet reduction, the word space is reduced from $5^L$ to $3^L$, and the time asymptotics of such an algorithm is naturally reduced to $\Theta(3^{L})$ from $\Theta(5^L)$.

In fact, this optimization allows us to extend the exhaustive search to $L=34$ and improve the result by almost $2$ orders of magnitude.

\subsubsection{Boundary-\texorpdfstring{$\sigma_3$}{sigma3} reduction}

This improvement is constant, not asymptotic. Since the only connecting operator is $\sigma_3$, and any other operator, being on the boundary, can be removed during the reduction to $[\CNOT]$ by the definition of Eq.~\ref{eq:CNOT_class}, it is meaningful to consider the matrix of operators that end and begin with $\sigma_3$. In this case, if we find a new record on length $L$ among three-generator braids with $\sigma_3$ at both ends of the braid, then we record this record, and if we do not find one, then we add any of the operators $\{\sigma_1, \sigma_2\}$ to either end of the braid --- this will not increase the distance to the class $[\CNOT]$, nor will it increase the leakage.

For a fixed length, there are $3^{L-2}$ inner words left instead of $3^L$. This is the same asymptotic behavior of $\Theta(3^L)$, but allows us to probe two consecutive lengths further without increasing the running time.

\subsubsection{Palindromic Reduction}

Palindromic elements of braid groups were systematically studied in ~\cite{DeloupPalindromic2008}. Among other things, they proved that the mapping $A\rightarrow AA^R$ is injective, where $A^R$ denotes the word $A$ read backwards, but not the inverse element $A^{-1}$. This means that elements of palindromic braids do not collapse into a braid of shorter length. By adding a single-letter center $M$, we obtain braids of odd length as well.

Similar structured and recursive braid constructions for Fibonacci anyons were used in~\cite{Carnahan2016}, but the positive-only family and its exhaustion criterion considered here are different.

All matrices $\sigma_i$ used are symmetric. Therefore,

\begin{equation}
    U(A^R)=U(A)^T,
    \qquad
    U\!\left(A M A^R\right)
    =
    U(A)^T U(M)U(A).
    \label{eq:palindromic_factorization}
\end{equation}

Therefore, for $L=2h+|M|$ it is sufficient to list only the halves of $A$ of length $h$. For a fixed finite set of centers, the number of constructions to be tested decreases from $\Theta(3^L)$ to

\begin{equation}
    \Theta(3^h)
    =
    \Theta\left(3^{L/2}\right)
    =
    \Theta\left((\sqrt{3})^L\right).
\end{equation}

It is this reduction that enables an exact search of the palindromic family up to $L=42$. However, it is not an equivalent transformation of the complete restricted space: the resulting optimum is guaranteed only within the $AMA^R$ family. We do not prove that all equivalent operators are expressible in terms of palindromes, but palindromes exist among the optimal braids for the preceding $L$ and, as can be seen from the graph Fig.~\ref{fig:curve}, have approximately the same asymptotic behavior as the previous constructions.

\subsubsection{Garside quotient of palindromic halves}

We group the two consecutive total lengths $L=2H-1$ and $L=2H$ into a single enumeration level $H=\lceil L/2\rceil$. In this subsection, we enumerate one canonical representative of each positive half-braid using Garside left normal form~\cite{Bokut2008Garside,GebhardtTawn2014}.

Different words over $\{\sigma_1,\sigma_2,\sigma_3\}$ can still define a single element of the positive braid monoid $B_4^+$ due to Artin's relations

\begin{equation}
    \sigma_1\sigma_2\sigma_1=\sigma_2\sigma_1\sigma_2,
    \qquad
    \sigma_2\sigma_3\sigma_2=\sigma_3\sigma_2\sigma_3,
    \qquad
    \sigma_1\sigma_3=\sigma_3\sigma_1.
    \label{eq:positive_b4_relations}
\end{equation}

For example, the words $121$ and $212$, as well as $13$ and $31$, should not be considered distinct and tested separately, since they correspond to the same matrix. Conversely, calculating various statistics and distribution functions for such Garside quotient braids may be distorted, since we are cutting out some states from the outcome space, and the statistics may be biased.

We enumerate not words, but directly the elements of $B_4^+$, represented in Garside left normal form \cite{Bokut2008Garside}. We generate only canonical representatives, rather than factoring the entire set, which allows us to improve the asymptotics. Since reversing the order of letters is an antiautomorphism of the monoid and preserves relations~\eqref{eq:positive_b4_relations}, replacing a half-braid by its canonical representative does not change the corresponding braid $AMA^R$. Therefore, quotient exactly removes word duplicates and does not exclude any candidate from the chosen palindromic family. The enumeration uses the finite automaton of Garside normal forms described in~\cite[Sections~4.2--4.3]{GebhardtTawn2014}. We count admissible continuations by their total atomic length, including the initial power of the Garside half-twist $\Delta$, and use these counts to reconstruct each canonical representative from its integer rank; disjoint rank intervals are processed independently on the GPU. Reconstruction from an integer index is also used in~\cite[Section~2]{GebhardtGonzalezMeneses2013}, but their representatives are lexicographically minimal, whereas our ranks refer to Garside normal forms.

Let $a_h$ denote the number of elements of $B_4^+$ of atomic length $h$, where the atomic length is the length in the standard generators $\sigma_1,\sigma_2,\sigma_3$. The generating function of the sequence $(a_h)_{h\geq0}$ is~\cite{FloresGonzalezMeneses2022Growth}

\begin{equation}
    \sum_{h\geq0}a_h t^h
    =
    \frac{1}{1-3t+t^2+2t^3-t^6}.
    \label{eq:b4_positive_growth}
\end{equation}
The pole closest to the origin is $r=0.4792046224\ldots$, so
\begin{equation}
    a_h\sim C\rho^h,
    \qquad
    \rho=r^{-1}=2.0867912228\ldots .
    \label{eq:b4_positive_asymptotic}
\end{equation}
For the even branch $L=2H$, we enumerate $A=\sigma_3Y$ with $|Y|=H-1$ and test $AA^R$. For the odd branch $L=2H-1$, we use $A=\sigma_3Y$ with $|Y|=H-2$ and test $A\sigma_cA^R$ separately for $c\in\{1,2,3\}$. Thus, the number of candidates at level $H$ is
\begin{equation}
    Q_H=a_{H-1}+3a_{H-2}=\Theta(\rho^H).
    \label{eq:quotient_candidate_count}
\end{equation}

Thus, the number of palindromic candidates to be tested decreases from
\begin{equation}
    \Theta\!\left(3^{L/2}\right)
    \quad\text{to}\quad
    \Theta\!\left(\rho^{L/2}\right)
    =
    \Theta\!\left((\sqrt{\rho})^L\right)
    \approx
    \Theta\!\left(1.44457^L\right).
    \label{eq:garside_reduction}
\end{equation}
This is an exact reduction of the already chosen palindromic space, not an additional restriction on the family. Quotient search is complete for $H=22,\ldots,36$, which corresponds to the total lengths $L=43,\ldots,72$ and completes the chain of search optimizations within this paper.

\subsection{Sampling Rejection}

Burke et al. paper~\cite{Burke2024} has two main parts: exhaustive search up to a certain length and sampling on long braids, for which exhaustive search is computationally too difficult. However, the efficiency of sampling turns out to decrease with increasing leakage requirements. Indeed, suppose we have established an experimental fact in which the dependence of the minimum leakage on braid length can be approximated by an empirical dependence:

\begin{equation}
\ell_{\mathrm{fit}}(L) = \sqrt{5}\cdot 10^{1-\frac{L}{6}}
\label{eq:l_fit}
\end{equation}

This means that when searching for the next exhaustive search record, we will cut off candidates using the threshold:

\begin{equation}
\ell_L = 4\cdot\sqrt{5}\cdot 10^{1-\frac{L}{6}} \Rightarrow
L = -6\log_{10}\ell_L+C
\label{eq:tau_exh}
\end{equation}

The best asymptotics of a complete search for leakage using only part of the optimizations associated with reducing the number of generators will be of the order of:

\begin{equation}
n^{\mathrm{exh}}_L = \Theta\left(\sqrt{3}^L\right) = \Theta\left(3^{\frac{-6\log_{10}\ell_L+C}{2}}\right) =
\Theta\left(3^{-3\frac{\log_{3}\ell_L}{\log_{3}10}}\right) = \Theta\left(\ell_L^{-3\log_{10}3}\right) \approx \Theta\left(\ell_L^{-1.431}\right)
\label{eq:n_exh}
\end{equation}

To estimate from below the number of samples required to fall near the low-leakage classes $[\CNOT]$, we first estimate the number of samples required to fall into the low-leakage set of operators that are not necessarily close to the class $[\CNOT]$ with probability greater than $\frac{1}{2}$. Based on the empirically measured and proven formula in the Appendix~\ref{app:leak_CDF} for arbitrary $d$, we obtain in the low-leakage approximation:

\begin{equation}
\label{eq:sampling}
\begin{gathered}
P(n^{\mathrm{sample}}_L)
=
\left(1-\Pr(\ell<\ell_L)\right)^{n^{\mathrm{sample}}_L} \approx
\left(1-F_\infty(\ell_L)\right)^{n^{\mathrm{sample}}_L}
\leq \frac{1}{2}
\\
\left(1-\ell_L^{d-1}\right)^{n^{\mathrm{sample}}_L}
\leq \frac{1}{2}
\\
n^{\mathrm{sample}}_L
\log_2\left(1-\ell_L^{d-1}\right)
\leq -1
\\
n^{\mathrm{sample}}_L
\geq
-\frac{1}{\log_2\left(1-\ell_L^{d-1}\right)} =
O\left(-\frac{1}{-\ell_L^{d-1}}\right) =
O\left(\ell_L^{1-d}\right)
\end{gathered}
\end{equation}

Thus, when comparing the lower bound on the number of operations required to solve a problem using sampling and the upper bound on the number of operations required for a complete search, we obtain a clear advantage of the second method.

In this case, sampling at long lengths to achieve low $\ell_L$ is justified only in the case of $d \leq 2$. In our case, for a five-generator search $d=5$, and for a $3$-generator search $d=3$. The advantage of sampling at short lengths, which was used by Burke et al.~\cite{Burke2024}, is explained by the much lower complexity constant in the asymptotics. This observation is also confirmed by our pilot experiments. In our case, the use of sampling is not justified, so we do not present it in this article and limit ourselves only to various optimizations and reductions of exhaustive methods.

\section{Conclusion}

In this paper, we construct an exhaustive GPU pipeline for searching two-qubit gates of the $[\CNOT]$ class for Fibonacci anyons on some predefined subclasses. A combination of MITM factorization, adaptive leakage filtering, and batch GEMM computations allowed us to extend the full five-generator search from $L=14$ to $L=21$. At this length, we found a candidate with $\ell=0.00394572$ and $d_{[\CNOT]}\simeq7.8\cdot10^{-9}$, whose leakage is less than that of the best sampling result of Burke et al.~\cite{Burke2024} at $L=39$.

Further progress was achieved by successively narrowing the search space. The transition to the alphabet $\{\sigma_1,\sigma_2,\sigma_3\}$, fixing the boundary $\sigma_3$ and the palindromic restriction $AMA^R$ made it possible to extend the exhaustive search within the corresponding families to $L=72$. The best candidate found in it for $L=69$ has
\begin{equation}
\ell=8.78396\cdot10^{-11},
\qquad
d_{[\CNOT]}=4.26\cdot10^{-10}.
\end{equation}
We also proved that leakage is quantized in $\mathbb Z[\varphi]$, and any nonzero value of it obeys an exponential lower bound in the braid length. We also found the distribution function for operators built on an incomplete braid basis:
\[
F_{d,\infty}(\ell)=\ell^{d-1}
\]
Combined with the observed decrease in the record threshold with increasing $L$, this allows a quantitative comparison of random sampling with exhaustive structured search and explains why, in the range studied, simply increasing the sampling budget is ineffective compared to reducing the search space.

Thus, the main result consists not only of obtaining new low-leakage braids, but also of constructing a sequence of precisely defined search spaces with explicit exhaustion boundaries. This decouples proven optima within given families from the heuristically chosen structure of the families themselves, preserving reproducibility and the precise meaning of each stated guarantee.

\section{Future work}

A direct continuation is the construction of a certified, restricted compiler for cabled two-qubit Mironov-Morozov gates~\cite{Cabling2025,MironovMorozov2026}. For explicitly specified $SU(N)_k$, an alphabet of cable exponents, and a cost constraint, we propose exhaustively constructing multi-sector leakage-phase Pareto fronts and determining the minimum lengths of CZ/CNOT candidates. This will allow us to move from individual numerical examples to systematic optimization with proven completeness within a given cable family, without claiming global optimality across the entire braid group.

The next task is to find the analytical structure of the discovered optima: repeating motifs, recurrent cable families, and lower bounds on the achievable leakage. Such analysis may also lead to an extension of the construction to higher representations and qudits. The methods of the present paper remain open to the mathematical explanation of the $5\to3$, boundary-$\sigma_3$ and palindromic constraint reductions, as well as to obtain finite-length mixing estimates for $F_L(\ell)$ in the joint $L\to\infty$, $\ell\to0$ regime.

\section*{Acknowledgments}

The author thanks Nikita Kryazhevskikh for beautiful discussions on the mathematical part of this work during a week this summer in Yekaterinburg. The research was carried out within the state assignment of Ministry of Science and Higher Education of the Russian Federation for IITP RAS.

\appendix

\section{The dimension of the Fibonacci anyon fusion space \label{app:fibonacci}}

As can be seen in Fig.~\ref{fig:bg-fusion}, the result of the fusion of two anyons can be a one-dimensional space of charge $\mathbf{1}$ and a one-dimensional space of charge $\tau$. For $\tau^{\times n}$, we can calculate the dimensions of the spaces of two topological charges recursively, expanding the chain to the sum $\mathbf1\dim V^{\mathbf1}_{\tau^{\times n}}+\tau \dim V^\tau_{\tau^{\times n}}$, merging all pairs of anyons obtained in the final chain. In this case, we can write the chain of transformations:

\begin{align*}
\tau^{\times (n + 1)} = \tau^{\times n} \times \tau = (\mathbf{1}\dim V^{\mathbf1}_{\tau^{\times n}}+\tau \dim V^\tau_{\tau^{\times n}})\times \tau = \tau \dim V^{\mathbf1}_{\tau^{\times n}} + \mathbf{1}\dim V^\tau_{\tau^{\times n}}+\tau \dim V^\tau_{\tau^{\times n}} = \\ = \mathbf{1}\dim V^\tau_{\tau^{\times n}}+\tau (\dim V^{\mathbf1}_{\tau^{\times n}} + \dim V^\tau_{\tau^{\times n}}) = \mathbf{1}\dim V^\mathbf{1}_{\tau^{\times (n+1)}} + \tau \dim V^\tau_{\tau^{\times (n+1)}} \Rightarrow
\\
\Rightarrow
\begin{cases}
\dim V^\mathbf{1}_{\tau^{\times (n+1)}} = \dim V^\tau_{\tau^{\times n}} \\
\dim V^\tau_{\tau^{\times (n+1)}} = \dim V^{\mathbf1}_{\tau^{\times n}} + \dim V^\tau_{\tau^{\times n}}
\end{cases}
\Rightarrow
\begin{cases}
\dim V^\tau_{\tau^{\times (n+1)}} = \dim V^\tau_{\tau^{\times n}} + \dim V^\tau_{\tau^{\times (n-1)}} \\
\dim V^\mathbf{1}_{\tau^{\times (n+1)}} = \dim V^\tau_{\tau^{\times (n-1)}} + \dim V^\tau_{\tau^{\times (n-2)}}  \\
\dim V^\mathbf{1}_{\tau^{\times (n+1)}} = \dim V^\tau_{\tau^{\times n}} \\
\end{cases}
\Rightarrow
\\
\Rightarrow
\begin{cases}
\dim V^\tau_{\tau^{\times (n+1)}} = \dim V^\tau_{\tau^{\times n}} + \dim V^\tau_{\tau^{\times (n-1)}} \\
\dim V^\mathbf{1}_{\tau^{\times (n+1)}} = \dim V^\tau_{\tau^{\times (n-1)}} + \dim V^\tau_{\tau^{\times (n-2)}}  \\
\dim V^\tau_{\tau^{\times n}} = \dim V^\mathbf{1}_{\tau^{\times (n+1)}} \\
\end{cases}
\Rightarrow
\begin{cases}
\dim V^\tau_{\tau^{\times (n+1)}} = \dim V^\tau_{\tau^{\times n}} + \dim V^\tau_{\tau^{\times (n-1)}} \\
\dim V^\mathbf{1}_{\tau^{\times (n+1)}} = \dim V^{\mathbf{1}}_{\tau^{\times n}} + \dim V^{\mathbf{1}}_{\tau^{\times (n-1)}}  \\
\end{cases}
\end{align*}

In this case, the dimensions of both spaces form independent Fibonacci sequences. This is where the name of these anyons comes from. We know that in the case $\dim V^\tau_{\tau} = 1$, $\dim V^\tau_{\tau^{\times 2}} = 1$, $\dim V^\tau_{\tau^{\times 3}} = 2$, these are the first terms of the series of dimensions of spaces of topological charge $\tau$, the initial conditions of the Fibonacci sequence. The exact formulas for the charges are $\dim V^\tau_{\tau^{\times n}}=F_n$ and $\dim V^{\mathbf1}_{\tau^{\times n}}=F_{n-1}$, they both have growth $\Theta(\varphi^n)$, where $\varphi$ is the quantum dimension of one anyon.

\section{Limiting distribution of leakage for positive braids}
\label{app:leak_CDF}

We will prove the limit laws of CDF for uniform enumeration of positive words in alphabets

\begin{equation}
\mathcal A_3=\{\sigma_1,\sigma_2,\sigma_3\},
\qquad
\mathcal A_5=\{\sigma_1,\sigma_2,\sigma_3,\sigma_4,\sigma_5\}.
\end{equation}

We use the invariant spaces

\begin{equation}
W_3=\operatorname{span}\{\NC,\ket{1\tau},\ket{\tau\tau}\},
\qquad W_5=\operatorname{span}\{\NC,\ket{11},\ket{1\tau},\ket{\tau1},\ket{\tau\tau}\}.
\end{equation}

Here $d=\dim W_d$, $q_3=3$, $q_5=5$: the space dimension and the number of generators play different roles. For a uniformly chosen word $w$ of length $L$, its matrix on $W_d$ has the form

\begin{equation}
U_{d,L}=U(w)|_{W_d}
=\left.(\sigma_{i_L}\cdots\sigma_{i_1})\right|_{W_d},
\qquad i_s\in\{1,\ldots,q_d\}.
\end{equation}

Let $\mathcal B(U(d))$ denote the Borel $\sigma$-algebra of the group $U(d)$, that is, the family of all its Borel subsets. For $B\in\mathcal B(U(d))$, the probability measure is determined by the proportion of words whose matrices fall in $B$:

\begin{equation}
\mu_{d,L}(B)
=\Pr\!\left(U_{d,L}\in B\right)
=\frac{\left|\{w\in\mathcal A_{q_d}^L:U(w)|_{W_d}\in B\}\right|}{q_d^L}.
\label{eq:finite-word-measure}
\end{equation}

Thus, $\mu_{d,L}$ is the matrix distribution, and $\mu_{d,L}(B)$ is a number. Different words with the same matrix are counted separately. For one step, we set $\mu_d=\mu_{d,1}$. The independence of the letters gives

\begin{equation}
\mu_d(B)=\frac1{q_d}\sum_{j=1}^{q_d}\delta_{\sigma_j|_{W_d}}(B),
\qquad
\mu_{d,L}(B)=(\mu_d^{*L})(B),
\end{equation}

where $\delta_V(B)$ is $1$ for $V\in B$ and $0$ otherwise, and $\mu_d^{*L}$ denotes the distribution of the product of $L$ independent one-step matrices. Therefore,

\begin{equation}
F_{q_d,L}(\ell_0)
=\mu_{d,L}\!\left(\{U\in U(d):\ell(U)<\ell_0\}\right).
\label{eq:finite-leak-cdf}
\end{equation}

As in Definition~\ref{def:CDF}, the threshold inequality is strict.

The group $PU(d)$ identifies matrices that differ in global phase. $\pi_d:U(d)\to PU(d)$ denotes the corresponding projection. From the explicit generators

\begin{equation}
\det(\sigma_j|_{W_3})=e^{2\pi i/5}\quad(1\leq j\leq3),
\qquad
\det\sigma_j=e^{\pi i/5}\quad(1\leq j\leq5).
\end{equation}

The determinant of the word therefore depends cyclically on $L$, which excludes the Haar limit on $U(d)$. However,

\begin{equation}
\ell(e^{i\theta}U)
=1-|e^{i\theta}U_{00}|^2
=1-|U_{00}|^2
=\ell(U),
\qquad\theta\in\mathbb R.
\label{eq:leak-projective}
\end{equation}

It is enough to study the projective distributions defined by the equality $\bar\mu_d(B)=\mu_d(\pi_d^{-1}(B))$ for $B\in\mathcal B(PU(d))$. The normalized Haar measure $m_{PU(d)}$ satisfies $m_{PU(d)}(PU(d))=1$ and $m_{PU(d)}(VB)\allowbreak=m_{PU(d)}(BV)\allowbreak=m_{PU(d)}(B)$ for each $V\in PU(d)$ and $B\in\mathcal B(PU(d))$. Weak convergence means $\int f\,d(\bar\mu_d^{*L})\to\int f\,dm_{PU(d)}$ as $L\to\infty$ for each continuous observable $f:PU(d)\to\mathbb R$. Portmanteau's theorem translates this convergence of means into convergence of probabilities: under weak convergence of probability measures, their values converge on every Borel set whose boundary has zero limit measure.

\begin{lemma}[Projective mixing criterion]
\label{lem:projective-mixing-criterion}

Let $d\geq2$, and let the support of the probability measure $\mu$ on $PU(d)$ generate a
dense subgroup. Then

\begin{equation}
\forall\,B\in\mathcal B(PU(d)):\quad
m_{PU(d)}(\partial B)=0
\ \Longrightarrow\
\lim_{L\to\infty}(\mu^{*L})(B)=m_{PU(d)}(B).
\end{equation}

Here $\partial B$ is the boundary of the set $B$.

\end{lemma}
\begin{proof}
The Kawada-Ito theorem~\cite{KawadaIto1940,Monakov2025} yields weak convergence to Haar measure on a compact group under two conditions: the support generates a dense subgroup and is not contained in the coset of a proper closed normal subgroup. The first condition is given. $PU(d)$ has a unique proper closed normal subgroup, $\{I\}$, so violating the second condition would mean a singleton support. However, the closure of a group generated by a single element is commutative and cannot coincide with $PU(d)$. The condition $m_{PU(d)}(\partial B)=0$ allows us to apply the above-stated corollary of Portmanteau's theorem and obtain the desired limit.
\end{proof}

To prove the density, we directly use the Fibonacci theory from Section $6$ of~\cite{FreedmanLarsenWang2002Density}, where the labels $0$ and $2$ correspond to the vacuum and $\tau$. Each anyon is represented by a hole in the disk labeled with its charge, and the outer boundary of the disk carries the total charge of all the anyons.

From the explicit matrix $\sigma_5$, it is clear that on $W_3$, the last two anyons have a total charge of $\tau$. Given the total vacuum charge, the first four also have a charge of $\tau$. Therefore, $W_3$ is the fusion space of four $\tau$ anyons with a total charge of $\tau$: the corresponding surface has five boundaries labeled $\tau$, including the outer one. The space $W_5$ describes six $\tau$ anyons with a total vacuum charge. The outer vacuum boundary can be sealed with a disk without changing the state space; what remains is a sphere with six holes, labeled $\tau$. Thus, these are the cases $(g,n)=(0,5)$ and $(0,6)$ from the specified section, where $g$ is the genus of the surface and $n$ is the number of boundaries with a non-vacuum label.

It remains to relate the transformations of these surfaces to braids. The transformation of a disk with holes is determined by the motion of the holes and their additional rotations around their own boundaries. The first is described by a braid, while the second, for fixed charges, acts by scalar phases that become trivial in $PU(d)$. Therefore, the projective action of a pure mapping class group is realized by pure braids: on four anyons for $W_3$ and on six for $W_5$.

\begin{corollary}[Mixing of two braid ensembles]
\label{lem:projective-mixing}

For the alphabets under consideration and the sets $B$ from
Lemma~\ref{lem:projective-mixing-criterion},

\begin{equation}
\lim_{L\to\infty}(\bar\mu_d^{*L})(B)=m_{PU(d)}(B),
\qquad d\in\{3,5\}.
\end{equation}
\end{corollary}
\begin{proof}

By Theorem 6.2 of \cite {FreedmanLarsenWang2002Density}, the projective
images of pure mapping class groups of the surfaces in question are dense
in $PU(3)$ and $PU(5)$, respectively. These images are realized by pure
braids that are included in the groups $B_4$ and $B_6$ under consideration.
Consequently,

\begin{equation}
\overline{\left\langle
\pi_d(\sigma_1|_{W_d}),\ldots,\pi_d(\sigma_{q_d}|_{W_d})
\right\rangle}=PU(d).
\label{eq:jones-projective-density}
\end{equation}

We apply the lemma~\ref{lem:projective-mixing-criterion}. Inverse generators are not required to be added: the criterion allows for asymmetric support.

\end{proof}

\begin{lemma}[Haar leakage distribution]
\label{lem:haar-leak-cdf}

Let $U$ be distributed according to Haar measure on $U(d)$, $d\geq2$, and $\ket v$ be a fixed unit vector. Then

\begin{equation}
\Pr\!\left(1-|\langle v,Uv\rangle|^2\leq\ell_0\right)
=\ell_0^{d-1},\qquad0\leq\ell_0\leq1.
\label{eq:haar-leak-law}
\end{equation}
\end{lemma}
\begin{proof}

By Haar invariance, $U\ket v$ is uniform on the complex unit sphere. It can be obtained by normalizing a vector with independent complex Gaussian coordinates $Z_j$ of density $\pi^{-1}e^{-|z|^2}$. Therefore, $|\langle v,Uv\rangle|^2$ is distributed as $|Z_1|^2/\sum_{j=1}^d|Z_j|^2$. The quantities $|Z_j|^2$ are independent and have density $e^{-t}$, $t\geq0$. For $0<\ell_0\leq1$, we obtain

\begin{equation}
\begin{aligned}
\Pr\!\left(1-\frac{|Z_1|^2}{\sum_{j=1}^d|Z_j|^2}\leq\ell_0\right)
&=\Pr\!\left(|Z_1|^2\geq\frac{1-\ell_0}{\ell_0}
\sum_{j=2}^d|Z_j|^2\right)\\
&=\prod_{j=2}^d\int_0^\infty e^{-t/\ell_0}\,dt
=\ell_0^{d-1}.
\end{aligned}
\end{equation}

When $\ell_0=0$, the probability is zero, since $\sum_{j=2}^d|Z_j|^2>0$ almost surely.

\end{proof}

\begin{theorem}[Limiting CDF of leakage]
\label{thm:positive-leak-cdf}

For every fixed $\ell_0\in[0,1]$, we have

\begin{equation}
\lim_{L\to\infty}F_{q_3,L}(\ell_0)=\ell_0^2,
\qquad
\lim_{L\to\infty}F_{q_5,L}(\ell_0)=\ell_0^4.
\end{equation}
\end{theorem}
\begin{proof}

The leakage is continuous and independent of the global phase, so it defines a continuous function on $PU(d)$. The image of the Haar measure on $U(d)$ under $\pi_d$ is equal to the Haar measure on $PU(d)$, therefore, the Haar distribution of the leakage is given by lemma~\ref{lem:haar-leak-cdf} and has no atoms. The boundary of the set $\{U\in PU(d):\ell(U)<\ell_0\}$ is contained in $\{U\in PU(d):\ell(U)=\ell_0\}$ and has Haar measure zero. Applying Corollary~\ref{lem:projective-mixing} to this set, we obtain $F_{q_d,L}(\ell_0)\to\ell_0^{d-1}$ for a fixed threshold. Substituting $d=3,5$ completes the proof.

\end{proof}


\begin{thebibliography}{99}
\bibitem{Grover1996} L.~K.~Grover, \textit{A fast quantum mechanical algorithm for database search}, Proc.\ 28th ACM STOC, 212 (1996). \url{https://arxiv.org/abs/quant-ph/9605043}

\bibitem{Shor1997} P.~W.~Shor, \textit{Polynomial-time algorithms for prime factorization and discrete logarithms on a quantum computer}, SIAM J.\ Comput.\ \textbf{26}, 1484 (1997). \url{https://arxiv.org/abs/quant-ph/9508027}

\bibitem{GidneyEkera2021} C.~Gidney, M.~Eker\aa, \textit{How to factor 2048 bit RSA integers in 8 hours using 20 million noisy qubits}, Quantum \textbf{5}, 433 (2021). \url{https://arxiv.org/abs/1905.09749}

\bibitem{Simon1997} D.~R.~Simon, \textit{On the power of quantum computation}, SIAM J.\ Comput.\ \textbf{26}, 1474 (1997). \url{https://doi.org/10.1137/S0097539796298637}

\bibitem{BernsteinVazirani1997} E.~Bernstein, U.~Vazirani, \textit{Quantum complexity theory}, SIAM J.\ Comput.\ \textbf{26}, 1411 (1997). \url{https://doi.org/10.1137/S0097539796300921}

\bibitem{Tang2019} E.~Tang, \textit{A quantum-inspired classical algorithm for recommendation systems}, Proc.\ 51st ACM STOC, 217 (2019). \url{https://arxiv.org/abs/1807.04271}

\bibitem{ChiaEtAl2020} N.-H.~Chia, A.~Gily\'en, T.~Li, H.-H.~Lin, E.~Tang, C.~Wang, \textit{Sampling-based sublinear low-rank matrix arithmetic framework for dequantizing quantum machine learning}, Proc.\ 52nd ACM STOC, 387 (2020). \url{https://arxiv.org/abs/1910.06151}

\bibitem{IsogaiMuraoYamasaki2026} N.~Isogai, M.~Murao, H.~Yamasaki, \textit{A quantum-inspired dequantization method for diagonally weighted matrix functions: Application to learning with optimized random features}, arXiv:2609.10729 (2026). \url{https://arxiv.org/abs/2609.10729}

\bibitem{Bennett1973} C.~H.~Bennett, \textit{Logical reversibility of computation}, IBM J.\ Res.\ Dev.\ \textbf{17}, 525 (1973). \url{https://doi.org/10.1147/rd.176.0525}

\bibitem{BeverlandEtAl2022} M.~E.~Beverland, P.~Murali, M.~Troyer, K.~M.~Svore, T.~Hoefler, V.~Kliuchnikov, G.~H.~Low, M.~Soeken, A.~Sundaram, A.~Vaschillo, \textit{Assessing requirements to scale to practical quantum advantage}, arXiv:2211.07629 (2022). \url{https://arxiv.org/abs/2211.07629}

\bibitem{Feynman1982} R.~P.~Feynman, \textit{Simulating physics with computers}, Int.\ J.\ Theor.\ Phys.\ \textbf{21}, 467 (1982). \url{https://doi.org/10.1007/BF02650179}

\bibitem{Lloyd1996} S.~Lloyd, \textit{Universal quantum simulators}, Science \textbf{273}, 1073 (1996). \url{https://doi.org/10.1126/science.273.5278.1073}

\bibitem{BerryAhokasCleveSanders2007} D.~W.~Berry, G.~Ahokas, R.~Cleve, B.~C.~Sanders, \textit{Efficient quantum algorithms for simulating sparse Hamiltonians}, Commun.\ Math.\ Phys.\ \textbf{270}, 359 (2007). \url{https://arxiv.org/abs/quant-ph/0508139}

\bibitem{LowChuang2019} G.~H.~Low, I.~L.~Chuang, \textit{Hamiltonian simulation by qubitization}, Quantum \textbf{3}, 163 (2019). \url{https://arxiv.org/abs/1610.06546}

\bibitem{DawsonNielsen2006} C.~M.~Dawson, M.~A.~Nielsen, \textit{The Solovay--Kitaev algorithm}, Quantum Inf.\ Comput.\ \textbf{6}, 81 (2006). \url{https://arxiv.org/abs/quant-ph/0505030}

\bibitem{BrylinskiBrylinski2002} J.-L.~Brylinski, R.~Brylinski, \textit{Universal quantum gates}, in \textit{Mathematics of Quantum Computation}, edited by R.~Brylinski and G.~Chen (Chapman \& Hall/CRC, 2002), p.~101. \url{https://arxiv.org/abs/quant-ph/0108062}

\bibitem{NielsenChuang2010}
M.~A.~Nielsen, I.~L.~Chuang,
\textit{Quantum Computation and Quantum Information},
10th Anniversary ed. (Cambridge University Press, Cambridge, 2010).
\url{https://doi.org/10.1017/CBO9780511976667}

\bibitem{Krantz2019}
P.~Krantz, M.~Kjaergaard, F.~Yan, T.~P.~Orlando, S.~Gustavsson,
W.~D.~Oliver,
\textit{A quantum engineer's guide to superconducting qubits},
Appl.\ Phys.\ Rev.\ \textbf{6}, 021318 (2019).
\url{https://doi.org/10.1063/1.5089550};
\url{https://arxiv.org/abs/1904.06560}

\bibitem{Hamming1950}
R.~W.~Hamming,
\textit{Error detecting and error correcting codes},
Bell Syst.\ Tech.\ J.\ \textbf{29}, 147 (1950).
\url{https://doi.org/10.1002/j.1538-7305.1950.tb00463.x}

\bibitem{vonNeumann1956}
J.~von Neumann,
\textit{Probabilistic logics and the synthesis of reliable organisms from
unreliable components},
in \textit{Automata Studies}, edited by C.~E.~Shannon and J.~McCarthy
(Princeton University Press, Princeton, 1956), pp.~43--98.
\url{https://doi.org/10.1515/9781400882618-003}

\bibitem{ShorQEC1995} P.~W.~Shor, \textit{Scheme for reducing decoherence in quantum computer memory}, Phys.\ Rev.\ A \textbf{52}, R2493 (1995). \url{https://doi.org/10.1103/PhysRevA.52.R2493}

\bibitem{AharonovBenOr2008} D.~Aharonov, M.~Ben-Or, \textit{Fault-tolerant quantum computation with constant error rate}, SIAM J.\ Comput.\ \textbf{38}, 1207 (2008). \url{https://arxiv.org/abs/quant-ph/9906129}

\bibitem{WoottersZurek1982} W.~K.~Wootters, W.~H.~Zurek, \textit{A single quantum cannot be cloned}, Nature \textbf{299}, 802 (1982). \url{https://doi.org/10.1038/299802a0}

\bibitem{DennisKitaevLandahlPreskill2002} E.~Dennis, A.~Kitaev, A.~Landahl, J.~Preskill, \textit{Topological quantum memory}, J.\ Math.\ Phys.\ \textbf{43}, 4452 (2002). \url{https://arxiv.org/abs/quant-ph/0110143}

\bibitem{FowlerEtAl2012} A.~G.~Fowler, M.~Mariantoni, J.~M.~Martinis, A.~N.~Cleland, \textit{Surface codes: Towards practical large-scale quantum computation}, Phys.\ Rev.\ A \textbf{86}, 032324 (2012). \url{https://arxiv.org/abs/1208.0928}

\bibitem{Litinski2019} D.~Litinski, \textit{A game of surface codes: Large-scale quantum computing with lattice surgery}, Quantum \textbf{3}, 128 (2019). \url{https://arxiv.org/abs/1808.02892}

\bibitem{BreuckmannEberhardt2021}
N.~P.~Breuckmann, J.~N.~Eberhardt,
\textit{Quantum low-density parity-check codes},
PRX Quantum \textbf{2}, 040101 (2021).
\url{https://doi.org/10.1103/PRXQuantum.2.040101};
\url{https://arxiv.org/abs/2103.06309}

\bibitem{BombinMartinDelgado2006}
H.~Bombin, M.~A.~Martin-Delgado,
\textit{Topological quantum distillation},
Phys.\ Rev.\ Lett.\ \textbf{97}, 180501 (2006).
\url{https://doi.org/10.1103/PhysRevLett.97.180501};
\url{https://arxiv.org/abs/quant-ph/0605138}

\bibitem{GKP2001}
D.~Gottesman, A.~Kitaev, J.~Preskill,
\textit{Encoding a qubit in an oscillator},
Phys.\ Rev.\ A \textbf{64}, 012310 (2001).
\url{https://doi.org/10.1103/PhysRevA.64.012310};
\url{https://arxiv.org/abs/quant-ph/0008040}

\bibitem{Kitaev2003} A.~Yu.~Kitaev,
\textit{Fault-tolerant quantum computation by anyons},
Ann.\ Phys.\ \textbf{303}, 2 (2003). \url{https://arxiv.org/abs/quant-ph/9707021}

\bibitem{NayakRMP2008} C.~Nayak, S.~H.~Simon, A.~Stern, M.~Freedman, S.~Das~Sarma,
\textit{Non-Abelian anyons and topological quantum computation},
Rev.\ Mod.\ Phys.\ \textbf{80}, 1083 (2008). \url{https://arxiv.org/abs/0707.1889}

\bibitem{BelovMorozovIsing2026}
A.~Belov, A.~Morozov,
\textit{Ising anyons in the $SU(2)_2$ Chern--Simons theory},
arXiv:2605.15268 (2026).
\url{https://arxiv.org/abs/2605.15268}

\bibitem{Bravyi2006} S.~Bravyi, \textit{Universal quantum computation with the $\nu=5/2$ fractional quantum Hall state}, Phys.\ Rev.\ A \textbf{73}, 042313 (2006). \url{https://arxiv.org/abs/quant-ph/0511178}

\bibitem{KolganovMorozov2020}
N.~Kolganov, A.~Morozov,
\textit{Quantum $\mathcal R$-matrices as universal qubit gates},
JETP Lett.\ \textbf{111}(9), 519--524 (2020).
\url{https://doi.org/10.1134/S0021364020090027};
\url{https://arxiv.org/abs/2004.07764}

\bibitem{KolganovMironovMorozov2023}
N.~Kolganov, S.~Mironov, A.~Morozov,
\textit{Large $k$ topological quantum computer},
Nucl.\ Phys.\ B \textbf{987}, 116072 (2023).
\href{https://doi.org/10.1016/j.nuclphysb.2023.116072}
{doi:10.1016/j.nuclphysb.2023.116072};
\url{https://arxiv.org/abs/2105.03980}.

\bibitem{BelovMorozov2024} A.~Belov, A.~Morozov,
\textit{Measuring Chern-Simons level $k$ by braiding $SU(2)_k$ anyons},
Eur.\ Phys.\ J.\ C \textbf{85}, 91 (2025). \url{https://arxiv.org/abs/2408.14188}

\bibitem{FreedmanLarsenWang2002Universal} M.~Freedman, M.~Larsen, Z.~Wang,
\textit{A modular functor which is universal for quantum computation},
Commun.\ Math.\ Phys.\ \textbf{227}, 605 (2002). \url{https://arxiv.org/abs/quant-ph/0001108}

\bibitem{MorozovTopologicalCharge2024}
A.~A.~Morozov,
\textit{On measuring the topological charge of anyons},
Probl.\ Inf.\ Transm.\ \textbf{60}(1), 28--34 (2024).
\url{https://doi.org/10.1134/S0032946024010046};
\url{https://arxiv.org/abs/2403.07847}

\bibitem{Cabling2025}
S.~Mironov, A.~Morozov,
\textit{Entangling gates from cabling of knots},
Eur.\ Phys.\ J.\ C \textbf{85}, 799 (2025). \url{https://arxiv.org/abs/2412.20931}

\bibitem{MironovMorozov2026}
S.~Mironov, A.~Morozov,
\textit{Entangling gates for the $SU(N)$ anyons},
arXiv:2605.04016 (2026). \url{https://arxiv.org/abs/2605.04016}

\bibitem{CuiWang2019} S.~X.~Cui, K.~T.~Tian, J.~F.~Vasquez, Z.~Wang, H.~M.~Wong,
\textit{The search for leakage-free entangling Fibonacci braiding gates},
J.\ Phys.\ A \textbf{52}, 455301 (2019). \url{https://arxiv.org/abs/1904.01731}

\bibitem{BremnerEtAl2002} M.~J.~Bremner, C.~M.~Dawson, J.~L.~Dodd, A.~Gilchrist, A.~W.~Harrow, D.~Mortimer, M.~A.~Nielsen, T.~J.~Osborne, \textit{Practical scheme for quantum computation with any two-qubit entangling gate}, Phys.\ Rev.\ Lett.\ \textbf{89}, 247902 (2002). \url{https://arxiv.org/abs/quant-ph/0207072}

\bibitem{JozsaLinden2003} R.~Jozsa, N.~Linden, \textit{On the role of entanglement in quantum-computational speed-up}, Proc.\ R.\ Soc.\ A \textbf{459}, 2011 (2003). \url{https://arxiv.org/abs/quant-ph/0201143}

\bibitem{Vidal2003} G.~Vidal, \textit{Efficient classical simulation of slightly entangled quantum computations}, Phys.\ Rev.\ Lett.\ \textbf{91}, 147902 (2003). \url{https://arxiv.org/abs/quant-ph/0301063}

\bibitem{Makhlin2002}
Yu.~Makhlin,
\textit{Nonlocal properties of two-qubit gates and mixed states, and the optimization of quantum computations},
Quantum Inf.\ Process.\ \textbf{1}, 243--252 (2002).
\href{https://doi.org/10.1023/A:1022144002391}
{doi:10.1023/A:1022144002391};
\url{https://arxiv.org/abs/quant-ph/0002045}.

\bibitem{BarencoEtAl1995} A.~Barenco, C.~H.~Bennett, R.~Cleve, D.~P.~DiVincenzo, N.~Margolus, P.~Shor, T.~Sleator, J.~A.~Smolin, H.~Weinfurter, \textit{Elementary gates for quantum computation}, Phys.\ Rev.\ A \textbf{52}, 3457 (1995). \url{https://arxiv.org/abs/quant-ph/9503016}

\bibitem{HormoziEtAl2007} L.~Hormozi, G.~Zikos, N.~E.~Bonesteel, S.~H.~Simon, \textit{Topological quantum compiling}, Phys.\ Rev.\ B \textbf{75}, 165310 (2007). \url{https://doi.org/10.1103/PhysRevB.75.165310}

\bibitem{KBS2014}
V.~Kliuchnikov, A.~Bocharov, K.~M.~Svore,
\textit{Asymptotically optimal topological quantum compiling},
Phys.\ Rev.\ Lett.\ \textbf{112}, 140504 (2014). \url{https://arxiv.org/abs/1310.4150}

\bibitem{XuWan2008}
H.~Xu, X.~Wan,
\textit{Constructing functional braids for low-leakage topological quantum computing},
Phys.\ Rev.\ A \textbf{78}, 042325 (2008). \url{https://arxiv.org/abs/0802.4213}

\bibitem{Carnahan2016}
C.~Carnahan, D.~Zeuch, N.~E.~Bonesteel,
\textit{Systematically generated two-qubit anyon braids},
Phys.\ Rev.\ A \textbf{93}, 052328 (2016). \url{https://arxiv.org/abs/1511.00719}

\bibitem{Long2025Fib}
J.~Long, X.~Huang, J.~Zhong, L.~Meng,
\textit{Genetic algorithm enhanced Solovay--Kitaev algorithm for quantum
compiling of Fibonacci anyons},
Phys.\ Scr.\ \textbf{100}, 115111 (2025).
\href{https://doi.org/10.1088/1402-4896/ae1d30}
{doi:10.1088/1402-4896/ae1d30};
\url{https://arxiv.org/abs/2501.01746}.

\bibitem{Metaplectic2025}
J.~Long, J.~Zhong, L.~Meng,
\textit{Topological quantum compilation of metaplectic anyons based on the genetic optimized algorithms},
Phys.\ Rev.\ A \textbf{112}, 022421 (2025). \url{https://arxiv.org/abs/2501.01745}

\bibitem{Long2026b}
J.~Long, Y.~Li, J.~Zhong, L.~Meng,
\textit{The construction of a universal quantum gate set for the $SU(2)_k$ ($k=5,6,7$) anyon models via genetic optimized algorithm},
Phys.\ Lett.\ A \textbf{565}, 131142 (2026). \url{https://arxiv.org/abs/2505.01774}

\bibitem{Long2026a}
J.~Long, Z.~Liu, Y.~Li, L.~Meng,
\textit{Universal quantum gate compilation in $SU(2)_k$ anyon models via multiple-braiding},
arXiv:2602.15324v4 (2026).
\url{https://arxiv.org/abs/2602.15324}.

\bibitem{Long2025NSI}
J.~Long, Z.~Liu, Y.~Li, J.~Zhong, L.~Meng,
\textit{High-fidelity universal quantum gate compilation for non-semisimple Ising anyons via genetic algorithm-optimized Solovay--Kitaev decomposition},
J.\ Phys.\ A \textbf{59}, 31LT01 (2026).
\url{https://doi.org/10.1088/1751-8121/ae91a0};
\url{https://arxiv.org/abs/2511.13194}.

\bibitem{Rytir2025}
P.~Ryt\'{\i}\v{r}, P.~C.~Burke, C.~Aravanis, J.~Vala, J.~Mare\v{c}ek,
\textit{Topological Quantum Compilation Using Mixed-Integer Programming},
IEEE Trans.\ Quantum Eng.\ \textbf{7}, 3104212 (2026).
\href{https://doi.org/10.1109/TQE.2026.3710775}
{doi:10.1109/TQE.2026.3710775};
\url{https://arxiv.org/abs/2511.09513}.

\bibitem{Burke2024}
P.~C.~Burke, C.~Aravanis, J.~Aspman, J.~Mare\v{c}ek, J.~Vala,
\textit{Topological quantum compilation of two-qubit gates},
Phys.\ Rev.\ A \textbf{110}, 052616 (2024). \url{https://arxiv.org/abs/2408.07132}

\bibitem{AmyMITM2013} M.~Amy, D.~Maslov, M.~Mosca, M.~Roetteler,
\textit{A meet-in-the-middle algorithm for fast synthesis of depth-optimal quantum circuits},
IEEE TCAD \textbf{32}, 818 (2013). \url{https://arxiv.org/abs/1206.0758}

\bibitem{Ruiz2025}
D.~Ruiz, J.~Guillaud, A.~Leverrier, M.~Mirrahimi, C.~Vuillot,
\textit{LDPC-cat codes for low-overhead quantum computing in 2D},
Nat.\ Commun.\ \textbf{16}, 1040 (2025).
\href{https://doi.org/10.1038/s41467-025-56298-8}
{doi:10.1038/s41467-025-56298-8}.

\bibitem{GidneyYoked2025}
C.~Gidney, M.~Newman, P.~Brooks, C.~Jones,
\textit{Yoked surface codes},
Nat.\ Commun.\ \textbf{16}, 4498 (2025).
\url{https://doi.org/10.1038/s41467-025-59714-1};
\url{https://arxiv.org/abs/2312.04522}

\bibitem{KlimovEtAl2024}
P.~V.~Klimov, A.~Bengtsson, C.~Quintana, et al.,
\textit{Optimizing quantum gates towards the scale of logical qubits},
Nat.\ Commun.\ \textbf{15}, 2442 (2024).
\url{https://doi.org/10.1038/s41467-024-46623-y};
\url{https://arxiv.org/abs/2308.02321}

\bibitem{LeinaasMyrheim1977}
J.~M.~Leinaas, J.~Myrheim,
\textit{On the theory of identical particles},
Nuovo Cimento B \textbf{37}, 1--23 (1977).
\href{https://doi.org/10.1007/BF02727953}
{doi:10.1007/BF02727953}.

\bibitem{RowellStongWang2009}
E.~Rowell, R.~Stong, Z.~Wang,
\textit{On Classification of Modular Tensor Categories},
Commun.\ Math.\ Phys.\ \textbf{292}, 343--389 (2009).
\href{https://doi.org/10.1007/s00220-009-0908-z}
{doi:10.1007/s00220-009-0908-z};
\url{https://arxiv.org/abs/0712.1377}.

\bibitem{ReadRezayi1999}
N.~Read, E.~Rezayi,
\textit{Beyond paired quantum Hall states: Parafermions and incompressible
states in the first excited Landau level},
Phys.\ Rev.\ B \textbf{59}, 8084--8092 (1999).
\href{https://doi.org/10.1103/PhysRevB.59.8084}
{doi:10.1103/PhysRevB.59.8084};
\url{https://arxiv.org/abs/cond-mat/9809384}.

\bibitem{Mong2017}
R.~S.~K.~Mong, M.~P.~Zaletel, F.~Pollmann, Z.~Papi\'c,
\textit{Fibonacci anyons and charge density order in the 12/5 and 13/5
quantum Hall plateaus},
Phys.\ Rev.\ B \textbf{95}, 115136 (2017).
\href{https://doi.org/10.1103/PhysRevB.95.115136}
{doi:10.1103/PhysRevB.95.115136}.

\bibitem{LevinWen2005}
M.~A.~Levin, X.-G.~Wen,
\textit{String-net condensation: A physical mechanism for topological phases},
Phys.\ Rev.\ B \textbf{71}, 045110 (2005).
\href{https://doi.org/10.1103/PhysRevB.71.045110}
{doi:10.1103/PhysRevB.71.045110};
\url{https://arxiv.org/abs/cond-mat/0404617}.

\bibitem{Artin1947}
E.~Artin,
\textit{Theory of Braids},
Ann.\ of Math.\ \textbf{48}, no.~1, 101--126 (1947).
\href{https://doi.org/10.2307/1969218}
{doi:10.2307/1969218}.

\bibitem{HadjiivanovGeorgiev2024} L.~Hadjiivanov, L.~S.~Georgiev,
\textit{Braiding Fibonacci anyons},
J.\ High Energy Phys.\ \textbf{2024}, 84 (2024). \url{https://arxiv.org/abs/2404.01778}

\bibitem{Jones1987Hecke}
V.~F.~R.~Jones,
\textit{Hecke algebra representations of braid groups and link polynomials},
Ann.\ of Math.\ \textbf{126}, no.~2, 335--388 (1987).
\href{https://doi.org/10.2307/1971403}
{doi:10.2307/1971403}.

\bibitem{AnokhinaMorozov2014Cabling}
A.~Anokhina, An.~Morozov,
\textit{Cabling procedure for the colored HOMFLY polynomials},
Theor.\ Math.\ Phys.\ \textbf{178}, 1--58 (2014).
\url{https://arxiv.org/abs/1307.2216}.

\bibitem{Suchara2015}
M.~Suchara, A.~W.~Cross, J.~M.~Gambetta,
\textit{Leakage suppression in the toric code},
Quantum Inf.\ Comput.\ \textbf{15}, 997--1016 (2015).
\href{https://doi.org/10.26421/QIC15.11-12-8}
{doi:10.26421/QIC15.11-12-8};
\url{https://arxiv.org/abs/1410.8562}.

\bibitem{AliferisTerhal2007}
P.~Aliferis, B.~M.~Terhal,
\textit{Fault-tolerant quantum computation for local leakage faults},
Quantum Inf.\ Comput.\ \textbf{7}, 139--156 (2007).
\href{https://doi.org/10.26421/QIC7.1-2-9}
{doi:10.26421/QIC7.1-2-9};
\url{https://arxiv.org/abs/quant-ph/0511065}.

\bibitem{AinsworthSlingerland2011}
R.~Ainsworth, J.~K.~Slingerland,
\textit{Topological qubit design and leakage},
New J.\ Phys.\ \textbf{13}, 065030 (2011).
\url{https://doi.org/10.1088/1367-2630/13/6/065030}

\bibitem{BigelowLevaillant2019}
S.~Bigelow, C.~Levaillant,
\textit{An exact entangling gate using Fibonacci anyons},
Bull.\ Aust.\ Math.\ Soc.\ \textbf{99}, 319--326 (2019).
\url{https://doi.org/10.1017/S0004972718001028}

\bibitem{FreedmanLarsenWang2002Density}
M.~H.~Freedman, M.~J.~Larsen, and Z.~Wang,
\newblock The two-eigenvalue problem and density of Jones representation
of braid groups,
\newblock \emph{Communications in Mathematical Physics}
\textbf{228} (2002), no.~1, 177--199.
\newblock
\href{https://doi.org/10.1007/s002200200636}
{doi:10.1007/s002200200636}.

\bibitem{Kuperberg2011} G.~Kuperberg,
\textit{Denseness and Zariski denseness of Jones braid representations},
Geom.\ Topol.\ \textbf{15}, 11--39 (2011). \url{https://arxiv.org/abs/0909.1881}

\bibitem{KawadaIto1940}
Y.~Kawada and K.~It\^o,
\newblock On the probability distribution on a compact group. I,
\newblock \emph{Proceedings of the Physico-Mathematical Society of Japan,
3rd Series}
\textbf{22} (1940), no.~12, 977--998.
\newblock
\href{https://doi.org/10.11429/ppmsj1919.22.12_977}
{doi:10.11429/ppmsj1919.22.12\_977}.

\bibitem{Mustafayev2022}
H.~Mustafayev,
\newblock A note on the Kawada--It\^o theorem,
\newblock \emph{Statistics \& Probability Letters}
\textbf{181} (2022), Article~109261.
\newblock
\href{https://doi.org/10.1016/j.spl.2021.109261}
{doi:10.1016/j.spl.2021.109261}.

\bibitem{Monakov2025}
G.~V.~Monakov,
\newblock Non-stationary It\^o--Kawada and ergodic theorems for random
isometries,
\newblock \emph{Groups, Geometry, and Dynamics} (2025), published online first.
\newblock
\href{https://doi.org/10.4171/GGD/856}
{doi:10.4171/GGD/856}.

\bibitem{Bhattacharya1972}
R.~N.~Bhattacharya,
\newblock Speed of convergence of the $n$-fold convolution of a probability
measure on a compact group,
\newblock \emph{Zeitschrift f\"ur Wahrscheinlichkeitstheorie und
Verwandte Gebiete}
\textbf{25} (1972), no.~1, 1--10.
\newblock
\href{https://doi.org/10.1007/BF00533331}
{doi:10.1007/BF00533331}.

\bibitem{DeloupPalindromic2008}
F.~Deloup, D.~Garber, S.~Kaplan, M.~Teicher,
\textit{Palindromic braids},
Asian J.\ Math.\ \textbf{12}, no.~1, 65--72 (2008).
\href{https://doi.org/10.4310/AJM.2008.v12.n1.a5}
{doi:10.4310/AJM.2008.v12.n1.a5};
\url{https://arxiv.org/abs/math/0410288}.

\bibitem{Bokut2008Garside}
L.~A.~Bokut,
\textit{Gröbner--Shirshov basis for the braid group in the
Artin--Garside generators},
J.\ Symbolic Comput.\ \textbf{43}, no.~6--7, 397--405 (2008).
\href{https://doi.org/10.1016/j.jsc.2007.02.003}
{doi:10.1016/j.jsc.2007.02.003};
\url{https://arxiv.org/abs/0806.1125}.

\bibitem{GebhardtTawn2014}
V.~Gebhardt, S.~Tawn,
\textit{Normal forms of random braids},
J.\ Algebra \textbf{408}, 115--137 (2014).
\href{https://doi.org/10.1016/j.jalgebra.2013.10.008}
{doi:10.1016/j.jalgebra.2013.10.008};
\url{https://arxiv.org/abs/1302.6676}.

\bibitem{GebhardtGonzalezMeneses2013}
V.~Gebhardt, J.~González-Meneses,
\textit{Generating random braids},
J.\ Combin.\ Theory Ser.\ A \textbf{120}, no.~1, 111--128 (2013).
\href{https://doi.org/10.1016/j.jcta.2012.07.003}
{doi:10.1016/j.jcta.2012.07.003};
\url{https://arxiv.org/abs/1112.5485}.

\bibitem{FloresGonzalezMeneses2022Growth}
R.~Flores, J.~González-Meneses,
\textit{On the growth of Artin--Tits monoids and the partial theta function},
J.\ Combin.\ Theory Ser.\ A \textbf{190}, 105623 (2022).
\href{https://doi.org/10.1016/j.jcta.2022.105623}
{doi:10.1016/j.jcta.2022.105623};
\url{https://arxiv.org/abs/1808.03066}.

\end{thebibliography}
\end{document}